\documentclass{article}

\usepackage{PRIMEarxiv}

\usepackage[utf8]{inputenc} 
\usepackage[T1]{fontenc}    
\usepackage{hyperref}       
\usepackage{url}            
\usepackage{booktabs}       
\usepackage{amsfonts}       
\usepackage{nicefrac}       
\usepackage{microtype}      
\usepackage{lipsum}
\usepackage{fancyhdr}       
\usepackage{graphicx}       
\graphicspath{{media/}}     

\usepackage{amsmath}
\usepackage{amssymb}
\usepackage{bm,bbm}
\usepackage{xcolor}

\title{$\rho$-CP: Open Source Dislocation Density Based Crystal Plasticity Framework for Simulating Temperature- and Strain Rate-Dependent Deformation
}

\author{
  Anirban Patra**, Suketa Chaudhary, Namit Pai, Tarakram Ramgopal, Sarthak Khandelwal, Adwitiya Rao \thanks{Current affiliation: Department of Materials Science and Engineering, University of Toronto, Canada} \\
  Department of Metallurgical Engineering and Materials Science
  \\
  Indian Institute of Technology Bombay \\
  Mumbai, MH 400076, India \\
  \texttt{**Corresponding author:anirbanpatra@iitb.ac.in} \\
   \And
  David L. McDowell \\
  School of Materials Science and Engineering, GWW School of Mechanical Engineering \\
  Georgia Institute of Technology \\
  Atlanta, GA 30332, USA \\
  \texttt{david.mcdowell@me.gatech.edu} \\
}

\begin{document}
\maketitle

\begin{abstract}
This work presents an open source, dislocation density based crystal plasticity modeling framework, $\rho$-CP. A Kocks-type thermally activated flow is used for accounting for the temperature and strain rate effects on the crystallographic shearing rate. Slip system-level mobile and immobile dislocation densities, as well slip system-level backstress, are used as internal state variables for representing the substructure evolution during plastic deformation. A fully implicit numerical integration scheme is presented for the time integration of the finite deformation plasticity model. The framework is implemented and integrated with the open source finite element solver, Multiphysics Object-Oriented Simulation Environment (MOOSE). Example applications of the model are demonstrated for predicting the anisotropic mechanical response of single and polycrystalline hcp magnesium, strain rate effects and cyclic deformation of polycrystalline fcc OFHC copper, and temperature and strain rate effects on the deformation of polycrystalline bcc tantalum. Simulations of realistic Voronoi-tessellated microstructures as well as Electron Back Scatter Diffraction (EBSD) microstructures are demonstrated to highlight the model's ability to predict large deformation and misorientation development during plastic deformation.
\end{abstract}


\section{Introduction}
\label{sec:intro}
Crystal plasticity modeling frameworks have been extensively used to study the microstructure-sensitive, anisotropic, elasto-plastic deformation of metallic systems \cite{mcdowell2008viscoplasticity, mcdowell2010perspective, roters2010overview}. Broadly speaking, these frameworks rely on the development of material-specific constitutive models of crystallographic deformation mechanisms responsible for dislocation mediated inelastic deformation at the grain and sub-grain level (see \cite{repetto1997micromechanical, kalidindi1998incorporation, kothari1998elasto, tjahjanto2008crystallographically, austin2011dislocation} for some representative examples). Implementation of these constitutive models in finite element frameworks allows the study of deformation in microstructures and structures, with the broad objectives of establishing structure-property correlations and their effect on the mechanical properties and performance. Specifically, crystal plasticity models have been used for studying texture evolution during processing and in-service conditions \cite{beyerlein2003modeling, li2005crystal, jia2012non, zhang2016virtual}, the effect of various microstructural attributes, such as grain orientations, inclusions and heterogeneities, on the local deformation behavior \cite{kysar2007high, zhang2010simulation, guery2016slip, guan2017crystal, ganesan2021effects}, orientation-dependent microscale and macroscale mechanical response \cite{zhang2012phenomenological, lim2014grain, bittencourt2019interpretation} and lifing predictions \cite{mcdowell2010microstructure, stopka2022simulated}. Note that crystal plasticity models have also been implemented in self consistent modeling frameworks \cite{lebensohn1993self, van2006multiscale} and Fast Fourier Transform (FFT)-based frameworks \cite{lebensohn2012elasto, roters2019damask}. Crystal plasticity models are generally considered to be the state-of-the-art and their use has increased significantly over the last couple of decades, leveraging the parallel implementation of finite element solvers and the wide availability of high performance computing resources.

While the finite deformation kinematics of plasticity in crystalline solids was fairly well established in the last century \cite{lee1969elastic, asaro1977strain, peirce1982analysis, asaro1985overview}, there has been significant research in last three decades on the development of constitutive equations for representing material- and microstructure-specific mechanisms of strengthening and substructure evolution in these frameworks. Power law based hardening models have been used extensively for representing slip system strengthening \cite{hutchinson1976bounds, asaro1985overview}, perhaps due to their simplicity and ease in estimating the associated material parameters. Physically-based models have been developed subsequently by assuming a Taylor-type hardening model due to dislocations \cite{taylor1934mechanism}, along with constitutive models for the statistically-representative evolution of dislocations during plastic deformation \cite{estrin1996dislocation, zikry1996inelastic, kocks2003physics, ma2004constitutive, roters2010overview}. In addition, consideration of twinning as a pseudo-slip deformation mode \cite{lebensohn1993self, kalidindi1998incorporation} has facilitated modeling of the associated shear mechanisms in face centered cubic and hexagonal close packed crystals, where deformation twinning is commonly observed at ambient and low temperatures, in addition to dislocation slip. Transformation-induced plasticity has also been considered in recent studies \cite{wong2016crystal, feng2022crystal}. Finally, it should also be mentioned that non-local crystal plasticity frameworks have been developed to model the effects of strain gradient plasticity on the size-dependent mechanical properties and microstructure evolution \cite{arsenlis1999crystallographic, gurtin2002gradient, evers2004non, mayeur2011dislocation, dunne2012crystal, pai2022study}. The reader is referred to \cite{mcdowell2010perspective, roters2010overview, yaghoobi2021crystal} for a detailed review of advances in the field of crystal plasticity modeling.

In recent years, there has been a concerted effort in the materials and mechanics community towards the development of open-source computational tools, which can be widely used by researchers. In this regard, several open-source modeling tools have been developed spanning the length and time scales of materials physics, from the atomistic scales to the meso- and macro-scales. For example, open-source tools exist for density functional theory calculations \cite{giannozzi2009quantum}, molecular dynamics \cite{thompson2022lammps}, discrete dislocation dynamics \cite{arsenlis2007enabling}, concurrent atomistic-continuum modeling \cite{xu2018pycac}, phase field \cite{tonks2012object, dewitt2020prisms, microsim}, crystal plasticity modeling \cite{huang1991user, roters2019damask, yaghoobi2019prisms}, and materials informatics \cite{brough2017materials}. Further, there are several open-source finite element solvers, which can be used for solving multi-physics problems and Partial Differential Equations (PDEs), in general \cite{permann2020moose, code_aster, alnaes2015fenics, jasak2007openfoam, MR3043640, badia2018fempar}. These tools are also complemented by several open-source pre- and post-processing tools \cite{stukowski2009visualization, quey2011large, groeber2014dream, ahrens2005paraview}, thus enabling a complete open-source eco-system for materials modeling. The work presented in this manuscript also represents a contribution in the same vein.

We present a physically-based crystal plasticity constitutive modeling framework that accounts for substructure evolution due to underlying mechanisms of dislocation strengthening, interaction and evolution during plastic deformation. The constitutive model (or its variant) has been previously used for studying orientation-dependent deformation and residual strain development in Zr alloys \cite{thool2020role}, process-induced residual strain development during additive manufacturing \cite{pokharel2019analysis}, irradiation hardening and plastic flow localization in ferritic-martensitic steels \cite{patra2012crystal, patra2015void, patra2016crystal}, and orientation- and temperature-dependent yield stress prediction due to non-Schmid stresses in bcc-Fe \cite{patra2014constitutive} and single crystal Ni-based superalloys \cite{ranjan2021crystal}. While these former studies were material-specific and implemented in the form of Fortran subroutines, we present a more general C++ based implementation of the constitutive model in this work, in order to facilitate the user to run crystal plasticity finite element simulations for the desired application, with minimum code development or implementation. We first present the constitutive model and an algorithm for the fully implicit time step integration of the same, along with its interface with the open-source finite element solver, Multiphysics Object-Oriented Simulation Environment (MOOSE) \cite{permann2020moose}. The application of the model is demonstrated with several examples. First, we predict the mechanical response of hexagonal closed packed (hcp) magnesium single and polycrystals deformed in plane strain compression. We then demonstrate application of the model to predict the strain rate-dependent compression response and the cyclic response of face centered cubic (fcc) copper polycrystals. We also use the model to predict the strain rate- and temperature-dependent deformation of body centered cubic (bcc) tantalum under a variety of loading conditions. Finally, we demonstrate the ability of the model to simulate experimentally measured Electron Back Scatter Diffraction (EBSD) microstructures of tantalum oligocrystals and predict misorientation development during deformation. 

The C++ source codes for the numerical implementation of this framework, along with the necessary input files for running the example simulations, are shared in the github repository: https://github.com/apatra6/rhocp

\section{Crystal Plasticity Framework}
\label{sec:constitutive}
The crystal plasticity model is formulated in the context of finite deformation kinematics, which naturally allows the consideration for large deformation plasticity. Physically based models for slip and twinning are used to account for plastic deformation. These include: (a) a thermally-activated flow rule for dislocation slip, which accounts for temperature- and rate-dependent effects on the crystallographic shearing rate, (b) dislocation density-based strengthening of slip systems during plastic deformation, (c) substructure evolution in terms of the slip system-level mobile and immobile dislocation densities, and (d) physically-based evolution of the slip system-level backstress that may contribute to intragranular directional hardening during cyclic deformation, manifested in the form of Bauschinger effect, for example. Constitutive equations related to these individual mechanisms are presented in this Section, while their numerical implementation is discussed in the following Sections.

\subsection{Finite Deformation Kinematics}
This finite deformation framework is based on the multiplicative decomposition of the deformation gradient, $\bm{F}$, into the elastic, $\bm{F^{e}}$, and plastic parts, $\bm{F^{p}}$ \cite{asaro1977strain}, i.e.,
\begin{equation}
    \bm{F} = \bm{F^{e}} \cdot \bm{F^{p}} 
\end{equation}
where $\bm{F^{p}}$ accounts for the plastic deformation from the reference (undeformed) configuration to an intermediate isoclinic configuration, and $\bm{F^{e}}$ accounts for the elastic deformation and rigid body rotation from the intermediate configuration to the current (deformed) configuration. $\bm{F^{p}}$ is related to the plastic part of the spatial velocity gradient, $\bm{L^{p}}$, as
\begin{equation}
    \dot{\bm{F^{p}}}=\bm{L^{p}} \cdot \bm{F^{p}}
\end{equation}
Further, $\bm{L^{p}}$ is given as the tensor sum of the crystallographic shearing rates over all possible slip systems, $N_s$, i.e.,
\begin{equation}
    \bm{L^{p}} = \sum_{\alpha=1}^{N_s}{\dot\gamma^{\alpha} \bm{m_{0}^{\alpha}} \otimes \bm{n_{0}^{\alpha}} } 
\end{equation}
where $\dot\gamma^{\alpha}$ is the crystallographic shearing rate due to slip on slip system $\alpha$, and $\bm{m_{0}^{\alpha}}$ and $\bm{n_{0}^{\alpha}}$ are the unit vectors along slip and slip plane normal directions in the reference configuration, respectively. $\dot\gamma^{\alpha}$ evolves as a function of the resolved shear stress, $\tau^{\alpha}$, and the internal state variables. In this framework, the substructure evolution during plastic deformation is assumed to be represented by three slip system-level internal state variables: mobile dislocation density, $\rho_{m}^{\alpha}$, immobile dislocation density, $\rho_{i}^{\alpha}$, and the slip system-level backstress, $\chi^{\alpha}$.

Generally speaking, twinning may be an additional mode of plastic deformation for certain materials, for example, in cubic crystals with low stacking fault energies and in low symmetry hcp crystals, where dislocation slip is not viable for certain loading orientations. In order to account for this, the plastic spatial velocity gradient, $\bm{L^{p}}$, may be modified to have additional terms by considering twinning as a pseudo-slip deformation mode \cite{kalidindi1998incorporation}, i.e.,
\begin{equation}
    \bm{L^{p}} = \sum_{\alpha=1}^{N_s}{\dot\gamma^{\alpha} \bm{m_{0}^{\alpha}} \otimes \bm{n_{0}^{\alpha}}} + \sum_{\beta=1}^{N_t}{\dot\gamma^{\beta} \bm{m_{0}^{\beta}} \otimes \bm{n_{0}^{\beta}}} 
\end{equation}
where $\dot\gamma^{\beta}$ is the crystallographic shearing rate due to twinning on deformation system $\beta$, $N_t$ is the number of twinning (pseudo-slip) systems, and $\bm{m_{0}^{\beta}}$ and $\bm{n_{0}^{\beta}}$ are the unit vectors along pseudo-slip and pseudo-slip plane normal directions for the corresponding twinning systems, respectively. It should be noted that the plastic (or inelastic) velocity gradient may also be modified to account for dislocation climb associated mechanisms \cite{geers2014coupled, chaudhary2022crystal}.

\subsection{Elastic Deformation}
\label{sec:elasticity}
The elastic Green strain tensor in the intermediate configuration is given as:
\begin{equation}
    \bm{E^{e}} = \frac{1}{2}(\bm{{F^{e}}^{T}} \cdot \bm{F^{e}} - \bm{I}) 
\end{equation}
Further, the second Piola-Kirchhoff (PK) stress tensor is obtained using $\bm{S}=\bm{C_{0}}: \bm{E^{e}}$, where $\bm{C_{0}}$ is the fourth rank elastic stiffness tensor in the intermediate configuration. The Cauchy stress tensor is derived from the PK stress as: $\bm{S} = det(\bm{F^{e}}) \bm{{F^{e}}^{-1}} \cdot \bm{\sigma} \cdot \bm{F^{e-T}}$. Finally, the resolved shear stress acting on the slip system $\alpha$ is estimated using the Schmid law as:
\begin{equation}
    \tau^{\alpha} = \bm{m^{\alpha}} \cdot \bm{\sigma} \cdot \bm{n^{\alpha}} = \frac{1}{det(\bm{F^{e}})} \bm{m_{0}^{\alpha}} \cdot \bm{S} \cdot \bm{n_{0}^{\alpha}}
\end{equation}
Here, $\bm{m^{\alpha}}$ and $\bm{n^{\alpha}}$ denote the unit vectors along the slip and slip plane normal directions in the current (deformed) configuration, and can be related to the corresponding vectors in the reference configuration using: $\bm{m^{\alpha}} = \bm{F^{e}} \cdot \bm{m_{0}^{\alpha}}$ and $\bm{n^{\alpha}} = \bm{n_{0}^{\alpha} \cdot \bm{F^{e-1}}}$.

\subsection{Kinetics of Plastic Deformation}
Plastic deformation generally occurs due to dislocation glide along preferred directions on close-packed planes in crystalline solids. This is a temperature- and rate-dependent phenomenon. The flow rule for crystallographic shearing rate due to dislocation glide has been conventionally represented using power law, sine hyperbolic, or Arrhenius-type thermally activated model forms. The latter representation allows a more physically-based consideration for the rate kinetics of dislocation glide, along with thermal activation. In this regard, a Kocks-type thermally activated flow rule \cite{kocks1975thermodynamics} has been widely used in the literature. In the present work, we model the crystallographic shearing rate due to slip using a similar model as
\begin{equation}
\label{eqn:flow rule slip}
\dot{\gamma}^{\alpha}=\begin{cases}
          \dot{\gamma}^{\alpha}_{0s}exp\left[-\frac{\Delta F^{\alpha}}{k_{b}T}\left(1 - \left(\frac{\left|\tau^{\alpha} - \chi^{\alpha} \right|- s^{\alpha}_{a}}{s^{\alpha}_{t}}\right)^{p^{\alpha}}\right)^{q^{\alpha}}\right]sgn\left(\tau^{\alpha} - \chi^{\alpha}\right); \left|\tau^{\alpha} - \chi^{\alpha} \right| > s^{\alpha}_{a}; \alpha \in \text{slip} \\
          0; \text{otherwise}   \\
     \end{cases}
\end{equation}

Here, $\dot{\gamma}^{\alpha}_{0s}$ represents the reference strain rate associated with dislocation glide, $\Delta F^{\alpha}$ represents the activation energy for dislocation glide in the absence of external stress, $k_{b}$ is the Boltzmann constant, $T$ is the absolute temperature, and $p^{\alpha}$ and $q^{\alpha}$ are parameters associated with the shape of the enthalpy curve. $s^{\alpha}_{a}$ is the non-directional athermal slip resistance due to the long range stress fields of obstacles, such as dislocations, while $s^{\alpha}_{t}$ represents the thermal slip resistance due to the short range obstacles, such as solute atoms, that can be overcome by thermal vibrations. The driving force for dislocation glide on a slip system is of the form: $\tau^{\alpha} - \chi^{\alpha}$, where $\tau^{\alpha}$ is the aforementioned resolved shear stress, while $\chi^{\alpha}$ is the slip system-level backstress representative of directional hardening. For material systems where non-Schmid deformation is observed, additional contributions to the driving force may also be considered \cite{patra2014constitutive, ranjan2021crystal}. The signum function (represented by $sgn$) accounts for the direction of forward and backward slip due to positive and negative values of $\tau^{\alpha} - \chi^{\alpha}$, respectively.

Twinning, when present, has generally been modeled as a pseudo-slip plastic deformation mode \cite{kalidindi1998incorporation}. We consider the same here and represent the crystallographic shearing rate due to twinning using a phenomenological power law model as
\begin{equation}
\label{eqn:flow rule twinning}
\dot{\gamma}^{\alpha}=\begin{cases}
          \dot{\gamma}^{\alpha}_{0t}\left(\frac{\tau^{\alpha} - \tau^{\alpha}_{0t}}{D^{\alpha}}\right)^{m^{\alpha}}; \tau^{\alpha} > \tau^{\alpha}_{0t}; \alpha \in \text{twin}  \\
          0; \text{otherwise}   \\
     \end{cases}
\end{equation}
Here, $\dot{\gamma}^{\alpha}_{0t}$ represents the reference strain rate associated with twinning, $\tau^{\alpha}_{0t}$ represents the threshold resistance to twinning, $D^{\alpha}$ represents the frictional drag resistance, and $m^{\alpha}$ is the rate sensitivity exponent. We note that this is a rather simple representation of the crystallographic shearing rate due to twinning and several advanced constitutive models accounting for the twin nucleation and growth kinetics have been proposed \cite{meyers2001onset, beyerlein2008dislocation, oppedal2012effect, cheng2017crystal, abdolvand2020nucleation}.

\subsection{Strength Contributions}
The athermal slip resistance, $s^{\alpha}_{a}$, may have several contributions due to the intrinsic lattice resistance, grain size strengthening (Hall-Petch effect), and dislocation strengthening. The additive sum of these contributions reflects in $s^{\alpha}_{a}$ as
\begin{equation}
\label{eqn:Athermal Slip resistance} 
    s^{\alpha}_{a} = \tau^{\alpha}_{0s} + \frac{k^{\alpha}_{HP}}{\sqrt{d_{g}}} + k^{\alpha}_{\rho}G b^{\alpha} \sqrt{\sum_{\xi = 1}^{N_{s}} A^{\alpha \xi} \rho^{\xi}}
\end{equation}
where $\tau^{\alpha}_{0s}$ represents the intrinsic lattice resistance, $k^{\alpha}_{HP}$ represents the Hall-Petch coefficient associated with grain size strengthening \cite{hall1951deformation, petch1953cleavage}, $d_{g}$ represents the grain size, $k^{\alpha}_{\rho}$ represents the Taylor-type strength coefficient associated with dislocation strengthening \cite{taylor1934mechanism}, $G$ represents the shear modulus, $b^{\alpha}$ represents the Burgers vector magnitude, $A^{\alpha \xi}$ represents the matrix of slip system-level dislocation interaction coefficients between slip systems $\alpha$ and $\xi$, and $\rho^{\xi} = \rho^{\xi}_m + \rho^{\xi}_i$ is the aforementioned total dislocation density on slip system $\xi$. 

The thermal slip resistance, $s^{\alpha}_{t}$, may have contributions from the frictional resistance to dislocation glide, such as that due to solid solution strengthening \cite{fleischer1963substitutional, labusch1970statistical}, especially in alloy systems. In bcc crystals, the (high) intrinsic Peierls-Nabarro stress \cite{nabarro1997theoretical} may also contribute to $s^{\alpha}_{t}$. As a first order approximation, we have assumed that $s^{\alpha}_{t}$ does not evolve with plastic deformation.

Further, we also assume that the resistance to twinning due to $\tau^{\alpha}_{0t}$ and $D^{\alpha}$, when present, does not evolve during plastic deformation. Again, we note that more advanced constitutive description of twinning, for example, twin interactions with dislocations have been considered elsewhere \cite{beyerlein2008dislocation}. The purpose of this work is to introduce a generalized constitutive modeling framework, which can be adapted to the materials system and application by including the necessary strengthening mechanisms.

It should also be noted that twins are expected to reorient once a characteristic shear strain, $\gamma^{\alpha}_{tw}$, is reached within the twin. The associated lattice rotation tensor, $\bm{R^{tw}}$, is generally given by \cite{christian1995deformation, zhang2012phenomenological}
\begin{equation}
\label{twin_reorient}
    \bm{R^{tw}} = -\bm{I} + 2\bm{n^{tw}} \otimes \bm{n^{tw}}
\end{equation}
where $\bm{I}$ is the identity tensor and $\bm{n^{tw}}$ is the normal to the twin plane. While this description of twin reorientation kinematics is more physically appealing, it is generally associated with numerical convergence issues \cite{zhang2012phenomenological}. An alternate approach was proposed \cite{graff2007yielding}, where the twin resistance was assumed to harden exponentially once the characteristic shear strain is reached. Accordingly, $D^{\alpha}$ can be modified to a Voce hardening model as
\begin{equation}
\label{twin_hardening}
    D^{\alpha} = \begin{cases}
    D^{\alpha}_{0} + h^{\alpha}_{0}\left ( \left(\frac{\sum_{\alpha = 1}^{N_t} \gamma^{\alpha}}{\gamma^{tw}}\right)^{m^{\alpha}_{th}} - 1\right ); \sum_{\alpha = 1}^{N_t} \gamma^{\alpha} > \gamma^{\alpha}_{tw} \\
    D^{\alpha}_{0}; \gamma^{\alpha} \leq \gamma^{\alpha}_{tw}
    \end{cases}
\end{equation}
Here, $D^{\alpha}_0$ is the drag resistance prior to twin reorientation, $h^{\alpha}_{0}$ is the hardening coefficient and $m^{\alpha}_{th}$ is the associated hardening exponent. We have implemented both these constitutive models for representing twin reorientation in our framework and presented the results in later Sections.

\subsection{Substructure Evolution}
As mentioned earlier, the substructure evolution has been considered primarily in terms of two Internal State Variables (ISVs), namely, mobile dislocation density, $\rho^{\alpha}_m$, and immobile dislocation density, $\rho^{\alpha}_i$. The slip system-level backstress, $\chi^{\alpha}$, may also considered as an additional ISV for applications where simulating cyclic loading and Bauschinger effect is of interest. The equations are adopted from previous studies \cite{patra2012crystal, patra2014constitutive, pokharel2019analysis, thool2020role, ranjan2021crystal, chaudhary2022crystal}, where the application of these substructure evolution models has been demonstrated to study thermomechanical deformation in various materials systems.

The rates of evolution of the mobile and immobile dislocation densities are given as:
\begin{equation}
\label{eqn:mobile evolution} 
    \dot{\rho}^{\alpha}_{m} = \frac{k^{\alpha}_{M}}{b^{\alpha}}\sqrt{ \sum_{\xi=1}^{N_{s}} \rho^{\xi}}\left|\dot{\gamma}^{\alpha}\right| - \frac{2R^{\alpha}_{c}}{b^{\alpha}}\rho^{\alpha}_{m}\left|\dot{\gamma}^{\alpha}\right| - \frac{k^{\alpha}_{I}}{b^{\alpha}\lambda^{\alpha}}\left|\dot{\gamma}^{\alpha}\right|
\end{equation}
\begin{equation}
\label{eqn:immobile evolution} 
    \dot{\rho}^{\alpha}_{i} = \frac{k^{\alpha}_{I}}{b^{\alpha}\lambda^{\alpha}}\left|\dot{\gamma}^{\alpha}\right| - k^{\alpha}_{D}\rho^{\alpha}_{i}\left|\dot{\gamma}^{\alpha}\right|
\end{equation}
The first term on the RHS of Equation \ref{eqn:mobile evolution} represents the multiplication of mobile dislocations at pre-existing dislocation segments \cite{essmann1979annihilation}, while the second term represents the mutual annihilation of dislocation dipoles within a critical capture radius, $R^{\alpha}_{c}$.  Trapping of mobile dislocations at other dislocation segments is represented by the third term, where $\lambda^{\alpha} = 1/\sqrt{\sum_{\xi=1}^{N_{s}} \rho^{\xi}}$ represents the dislocation mean free path. Consequently, these trapped dislocations are rendered immobile, which is reflected in the first term of Equation \ref{eqn:immobile evolution}, while the last term represents the annihilation of immobile dislocations due to dynamic recovery processes. The associated material parameters, $k^{\alpha}_{M}$, $R^{\alpha}_{c}$, $k^{\alpha}_{I}$ and $k^{\alpha}_{D}$, may be obtained by fitting the predicted stress-strain response to the experimental hardening response. We note that depending on the materials system, cross-slip of screw dislocations may be an additional mechanism of dislocation evolution during plastic deformation. Constitutive models for cross-slip have been developed in the past \cite{patra2012crystal, castelluccio2017mesoscale} and can be integrated into this framework in future work.

We have modeled the backstress evolution as a function of the dislocation density using a self-hardening relation \cite{chaudhary2022crystal}, i.e.,
\begin{equation}
\label{eqn:backstress} 
    \dot \chi^{\alpha} = \left(k_{\chi 1}^{\alpha}G b^{\alpha} \sqrt{\rho^{\alpha}}sgn\left(\tau^{\alpha} - \chi^{\alpha}\right) - k_{\chi 2}^{\alpha}\chi^{\alpha}\right)  \left|\dot{\gamma}^{\alpha}\right|
\end{equation}
This constitutive model is inspired from \cite{shenoy2008microstructure} and is in the form of a non-linear Armstrong-Frederick kinematic hardening model \cite{armstrong1966mathematical}, which considers the development of backstress along the direction of net applied shear stress (first term) and also has a recall/recovery term (second term). $k_{\chi 1}^{\alpha}$ and $k_{\chi 2}^{\alpha}$ are the associated material parameters. Note that micromechanical constitutive models for dislocation substructure evolution during cyclic loading have also been proposed in recent studies \cite{castelluccio2017mesoscale, zirkle2021micromechanical}.

The above set of equations comprise all the constitutive equations implemented in the present framework to represent the plastic deformation and associated microstructure evolution of crystalline systems. Depending on the materials system or application, one or more of the above mechanisms may not be utilized.

\subsection{Numerical Integration}
The constitutive equations presented in the previous Section are highly stiff, non-linear differential equations, which are generally difficult to integrate. Accurate numerical integration is essential for implementation and interfacing with finite element codes, which may otherwise lead to convergence issues. In this Section, we present a fully implicit numerical algorithm for the time step integration of the crystal plasticity model. This algorithm is inspired from previous works \cite{mcginty2001multiscale, ling2005numerical, mcginty2006semi}, where different implicit and semi-implicit approaches for integration of crystal plasticity models have been discussed.

For a given deformation gradient, $\bm{F}$, at any time step, the numerical integration algorithm decomposes the total deformation gradient into the elastic and plastic parts using a Newton-Raphson algorithm that solves for the increment of crystallographic shearing rate, $\Delta\dot \gamma^{\alpha}$. This is accomplished by formulating a function, $f(\dot \gamma^{\alpha})$ \cite{ling2005numerical, mcginty2001multiscale, cuitino1993computational}, such that
\begin{equation}
\label{eqn:f_eqn} 
    f^{\alpha} = f(\dot \gamma^{\alpha}) = \begin{cases}
    \dot \gamma^{\alpha} - \dot{\gamma}^{\alpha}_{0s}exp\left[-\frac{\Delta F^{\alpha}}{k_{b}T}\left(1 - \left(\frac{\left|\tau^{\alpha} - \chi^{\alpha} \right|- s^{\alpha}_{a}}{s^{\alpha}_{t}}\right)^{p^{\alpha}}\right)^{q^{\alpha}}\right]sgn\left(\tau^{\alpha} - \chi^{\alpha}\right); \alpha \in \text{slip} \\
    \dot \gamma^{\alpha} - \dot{\gamma}^{\alpha}_{0t}\left(\frac{\tau^{\alpha} - \tau^{\alpha}_{0t}}{D^{\alpha}}\right)^{m^{\alpha}}; \alpha \in \text{twin} \\
    \end{cases}
\end{equation}
The above equation is obtained by rearranging terms in the respective flow rules for slip (Equation \ref{eqn:flow rule slip}) and twinning (Equation \ref{eqn:flow rule twinning}). Using the chain rule of differentiation, $f^{\alpha}_{i+1}$ can be written as:
\begin{equation}
    f^{\alpha}_{i+1} = f(\dot \gamma^{\alpha})_{i+1} = f(\dot \gamma^{\alpha})_{i} + \sum_{\beta=1}^{N_{s}+N_{t}} \frac{\partial{f(\dot \gamma^{\alpha})}}{\partial{\dot \gamma^{\beta}}}\Delta\dot \gamma^{\beta} 
\end{equation}
where the subscript, $i$, denotes the corresponding iteration number at any given time step. By iterative Newton-Raphson method, the function, $f^{\alpha}$, needs to be minimized, i.e., $f^{\alpha}_{i+1} \rightarrow 0$. Accordingly,
\begin{equation}
    f^{\alpha}_{i} = - \sum_{\beta=1}^{N_{s}+N_{t}} \frac{\partial{f^{\alpha}}}{\partial{\dot \gamma^{\beta}}}\Delta\dot \gamma^{\beta}    
\end{equation}
This procedure needs to be followed for all slip, $N_s$, and twin systems, $N_t$. By formulating a vector, $\bm{f} = [f^{\alpha}]; \alpha \in [1,(N_{s} + N_{t})]$, over all possible slip and twin systems and minimizing this vector, the converged values of $\bm{\Delta \dot \gamma} = [\Delta\dot \gamma^{\alpha}]; \alpha \in [1,(N_{s} + N_{t})]$ can thus be simultaneously obtained at any given time step by inverting the above expression. In the index notation, this can be written as
\begin{equation}
    \Delta\dot \gamma^{\alpha} = - \left[ {\frac{\partial{f^{\beta}}}{\partial{\dot \gamma^{\alpha}}}} \right] ^{-1}f^{\beta}; \alpha, \beta \in [1,(N_{s} + N_{t})]
\end{equation}
The main challenge lies in computing the partial derivatives associated with the above expression and is described in the following. 

By differentiating Equation \ref{eqn:f_eqn} with respect to $\dot \gamma^{\beta}$, we have
\begin{equation}
\frac{\partial{f^{\alpha}}}{\partial{\dot \gamma^{\beta}}} =
\begin{cases}
    \delta^{\alpha \beta} - \dot \gamma^{\alpha}\left[q^{\alpha} \frac{\Delta F^{\alpha}}{k_{b} T}\left(1-\left(\frac{\left|\tau^{\alpha} - \chi^{\alpha} \right| - s^{\alpha}_a}{s^{\alpha}_t}\right)^{p^{\alpha}}\right)^{q^{\alpha}-1}\right] \cdot
    \left[p^{\alpha}\left(\frac{\left|\tau^{\alpha} - \chi^{\alpha} \right| - s^{\alpha}_a}{s^{\alpha}_t}\right)^{p^{\alpha}-1}\right] \cdot \\
    \left[\frac{1}{s^{\alpha}_t}\left( \left(\frac{\partial\tau^{\alpha}}{\partial{\dot \gamma^{\beta}}} - \frac{\partial \chi^{\alpha}}{\partial{\dot \gamma^{\beta}}} \right)sgn\left(\tau^{\alpha} - \chi^{\alpha}\right) - \frac{\partial{s^{\alpha}_a}}{\partial{\dot \gamma^{\beta}}}\right) - \left(\frac{\left|\tau^{\alpha} - \chi^{\alpha} \right| - s^{\alpha}_a}{s^{\alpha 2}_t}\right)\frac{\partial{s^{\alpha}_t}}{\partial{\dot \gamma^{\beta }}}\right]; \alpha \in slip \\
    \delta^{\alpha \beta} - \frac{m^{\alpha} \dot \gamma^{\alpha}}{\tau^{\alpha} - \tau^{\alpha}_{0t}} \frac{\partial \tau^{\alpha}}{\partial \dot \gamma^{\beta}}; \alpha \in twin \\
\end{cases}
\end{equation}
Further, the individual partial derivatives are given as:
\begin{equation}
    \frac{\partial s^{\alpha}_{a}}{\partial \dot \gamma^{\beta}} = \frac{k^{\alpha}_{\rho}G b^{\alpha}}{2 \sqrt{\sum_{\xi = 1}^{N_{s}} A^{\alpha \xi} \rho^{\xi}}} \sum_{\xi = 1}^{N_{s}} A^{\alpha \xi} \left(\frac {\partial \rho^{\xi}_m}{\partial \dot \gamma^{\beta}} + \frac {\partial \rho^{\xi}_i}{\partial \dot \gamma^{\beta}} \right); \alpha \in slip
\end{equation}
The partial derivatives of the mobile and immobile dislocation density with respect to the crystallographic shearing rate are described later. Since the thermal slip resistance is assumed to be constant, its derivative is zero, i.e.,
\begin{equation}
    \frac{\partial s^{\alpha}_{t}}{\partial \dot \gamma^{\beta}} = 0; \alpha \in slip
\end{equation}
From \cite{mcginty2001multiscale}, the derivative of the resolved shear stress with respect to the crystallographic shearing rate may be approximated as:
\begin{equation}
    \frac{\partial \tau^{\alpha}}{\partial \dot \gamma^{\beta}} \approx - (\bm{m_{0}^{\alpha}} \otimes \bm{n_{0}^{\alpha}}) : \bm{C_{0}} : (\bm{m_{0}^{\alpha}} \otimes \bm{n_{0}^{\beta}})
\end{equation}
Partial derivative of the slip system-level back stress with respect to the shearing rate is computed in the following steps:
\begin{equation}
\begin{split}
    \frac{\partial \chi^{\alpha}}{\partial \dot \gamma^{\beta}} \approx \frac{\partial \dot \chi^{\alpha}}{\partial \dot \gamma^{\beta}} \Delta t = \\
    \left[\frac{k_{\chi 1}^{\alpha}G b^{\alpha}}{2 \sqrt{\rho^{\alpha}}} \left(\frac {\partial \rho^{\alpha}_m}{\partial \dot \gamma^{\beta}} + \frac {\partial \rho^{\alpha}_i}{\partial \dot \gamma^{\beta}} \right) sgn\left(\tau^{\alpha} - \chi^{\alpha} \right) \left|\dot{\gamma}^{\alpha}\right|
    + k_{\chi 1}^{\alpha}G b^{\alpha} \sqrt{\rho^{\alpha}} \delta^{\alpha \beta} - k_{\chi 2}^{\alpha} \frac{\partial \chi^{\alpha}}{\partial \dot \gamma^{\beta}} \left|\dot{\gamma}^{\alpha}\right| - k_{\chi 2}^{\alpha} \chi^{\alpha} \delta^{\alpha \beta} sgn\left(\dot \gamma^{\alpha} \right) \right] \Delta t; \\
    \alpha \in slip
\end{split}
\end{equation}
where $\Delta t$ is the time step increment. Rearranging terms,
\begin{equation}
    \frac{\partial \chi^{\alpha}}{\partial \dot \gamma^{\beta}} = \frac {\left(\frac{k_{\chi 1}^{\alpha}G b^{\alpha}}{2 \sqrt{\rho^{\alpha}}} \left(\frac {\partial \rho^{\alpha}_m}{\partial \dot \gamma^{\beta}} + \frac {\partial \rho^{\alpha}_i}{\partial \dot \gamma^{\beta}} \right) sgn\left(\tau^{\alpha} - \chi^{\alpha} \right) \left|\dot{\gamma}^{\alpha}\right|
    + k_{\chi 1}^{\alpha}G b^{\alpha} \sqrt{\rho^{\alpha}} \delta^{\alpha \beta} - k_{\chi 2}^{\alpha} \chi^{\alpha} \delta^{\alpha \beta} sgn\left(\dot \gamma^{\alpha} \right) \right) \Delta t}{1 + k_{\chi 2}^{\alpha} \left|\dot{\gamma}^{\alpha}\right| \Delta t}; \alpha \in slip
\end{equation}
The partial derivatives of the mobile and immobile dislocation densities with respect to $\dot \gamma^{\beta}$ also have to computed in multiple steps. These are described in the following.
\begin{equation}
\footnotesize
\begin{split}
    \frac{\partial \rho^{\alpha}_{m}}{\partial \dot \gamma^{\beta}} \approx \frac{\partial \dot \rho^{\alpha}_{m}}{\partial \dot \gamma^{\beta}} \Delta t = \\
    \left[\frac{k^{\alpha}_{M}}{2b\sqrt{\sum_{\xi=1}^{N_{s}} \rho^{\xi}}} \sum_{\xi=1}^{N_{s}} \frac{\partial \rho^{\xi}}{\partial \dot \gamma^{\beta}} -\frac{2R^{\alpha}_{c}}{b^{\alpha}} \frac{\partial \rho^{\alpha}_{m}}{\partial \dot \gamma^{\beta}} + \frac{k^{\alpha}_{I}}{b^{\alpha}\lambda^{\alpha 2}} \frac{\partial \lambda^{\alpha}}{\partial \dot \gamma^{\beta}} \right] \left|\dot{\gamma}^{\alpha}\right| \Delta t
    + \left[ \frac{k^{\alpha}_{M}}{b^{\alpha}} \sqrt{\sum_{\xi=1}^{N_{s}} \rho^{\xi}} - \frac{2R^{\alpha}_{c}}{b^{\alpha}} \rho^{\alpha}_{m} - \frac{k^{\alpha}_{I}}{b^{\alpha}\lambda^{\alpha}} \right] \delta^{\alpha \beta} sgn\left(\dot \gamma^{\alpha} \right) \Delta t
\end{split}    
\end{equation}
Here, the partial derivative of the dislocation mean free path is given as:
\begin{equation}
    \frac{\partial \lambda^{\alpha}}{\partial \dot \gamma^{\beta}} = \frac{\partial \lambda^{\alpha}}{\partial \rho^{\alpha}_{m}} \frac{\partial \rho^{\alpha}_{m}}{\partial \dot \gamma^{\beta}} \approx - \frac{\lambda^{\alpha 3}}{2} \frac{\partial \rho^{\alpha}_{m}}{\partial \dot \gamma^{\beta}}
\end{equation}
Using
\begin{equation}
    \frac{\partial \rho^{\alpha}_{m}}{\partial \dot \gamma^{\beta}} = \sum_{\xi=1}^{N_{s}} \delta^{\alpha \xi} \frac{\partial \rho^{\xi}_{m}}{\partial \dot \gamma^{\beta}}
\end{equation}
and rearranging terms, we arrive at
\begin{equation}
\begin{split}
    \frac{\partial \rho^{\xi}_{m}}{\partial \dot \gamma^{\beta}} = \left[\delta^{\xi \alpha} - \delta^{\xi \alpha} \frac{k^{\alpha}_{M}}{2b^{\alpha}\sqrt{\sum_{\xi=1}^{N_{s}} \rho^{\xi}}} \left|\dot{\gamma}^{\alpha} \right| \Delta t + \delta^{\xi \alpha} \frac{2R^{\alpha}_{c}}{b^{\alpha}} \left|\dot{\gamma}^{\alpha}\right| \Delta t + \delta^{\xi \alpha} \frac{k^{\alpha}_{I} \lambda^{\alpha}}{2 b^{\alpha}} \left|\dot{\gamma}^{\alpha}\right| \Delta t \right]^{-1} \cdot \\
    \left[\frac{k^{\alpha}_{M}}{b^{\alpha}} \sqrt{\sum_{\xi=1}^{N_{s}} \rho^{\xi}} \delta^{\alpha \beta} - \frac{2R^{\alpha}_{c}}{b^{\alpha}} \rho^{\alpha}_{m} \delta^{\alpha \beta} - \frac{k^{\alpha}_{I}}{b^{\alpha}\lambda^{\alpha}} \delta^{\alpha \beta} \right] sgn\left(\dot \gamma^{\alpha} \right) \Delta t; \alpha \in slip
\end{split}    
\end{equation}
Following a similar procedure, the partial derivative of immobile dislocation density with respect to the crystallographic shearing rate is
\begin{equation}
    \frac{\partial \rho^{\alpha}_{i}}{\partial \dot \gamma^{\beta}} = \frac{\left[\frac{k^{\alpha}_{I}}{b^{\alpha}\lambda^{\alpha}} \delta^{\alpha \beta} - k^{\alpha}_{D} \rho^{\alpha}_{i} \delta^{\alpha \beta} \right] sgn\left(\dot \gamma^{\alpha} \right) \Delta t}{1 - \frac{k^{\alpha}_{I} \lambda^{\alpha}}{2 b^{\alpha}} \left|\dot{\gamma}^{\alpha}\right| \Delta t + k^{\alpha}_{D} \left|\dot{\gamma}^{\alpha}\right| \Delta t}
\end{equation}
This implicit Newton-Raphson algorithm has been implemented together with a time step sub-incrementation algorithm \cite{mcginty2001multiscale} for accelerated convergence. Further, we have used a weighted convergence criterion \cite{mcginty2001multiscale}, in which the convergence of the $i^\text{th}$ iteration is determined by a weighted residual, $r$, as
\begin{equation}
    r = \frac{1}{N_{s} + N_{t}} \sqrt{\sum_{\xi=1}^{N_{s} + N_{t}} \frac{|\dot \gamma^{\alpha}|}{\dot \gamma_{max}} f^{\alpha 2}_{i}} \le tolerance
\end{equation}
where a user-defined tolerance can be specified depending on the imposed strain rate. The residual, $r$, is essentially the root mean squared error, weighted by the ratio of the absolute crystallographic shearing rate on a given deformation system, $\alpha$, to the maximum value, $\dot \gamma_{max} = max(| \dot \gamma^{\alpha} |), \alpha \in slip, twin$, and summed over all slip, $N_s$, and twin systems, $N_t$. As discussed in \cite{mcginty2001multiscale}, this weighted convergence criterion may reduce the convergence time by up to an order of magnitude, as compared to an unweighted convergence criterion. Note that terms related to twinning may be absent in examples where pseudo-slip systems due to twinning are not needed.

The converged values of stress, crystallographic shearing rates and internal state variables are used to compute the tangent stiffness tensor according to the following relation:
\begin{equation}
    \frac{\partial \bm \dot \sigma}{\partial \bm D} = \left[\bm C^{-1}_{0} + \frac{\partial \bm D^{p}}{\partial \bm \sigma} \Delta t \right] ^{-1}
\end{equation}
where $\frac{\partial \bm D^{p}}{\partial \bm \sigma}$ is given as \cite{mcginty2001multiscale}
\begin{equation}
    \frac{\partial \bm D^{p}}{\partial \bm \sigma} = \sum_{\alpha=1}^{N_s + N_t} \frac{\partial \bm D^{p}}{\partial \dot \gamma^{\alpha}} \frac{\partial \dot \gamma^{\alpha}}{\partial \tau^{\alpha}} \frac{\partial \tau^{\alpha}}{\partial \bm \sigma}
\end{equation}
The first and last terms on the RHS of the above equation are related to the Schmid tensor, $\bm m^{\alpha} \otimes \bm n^{\alpha}$, while the second term can be easily derived from the corresponding flow rules for dislocation glide and twinning.

These constitutive equations and their numerical implementation may be modified appropriately to account for additional deformation and strengthening mechanisms within the same crystal plasticity framework, as necessary. The reader may refer to \cite{chaudhary2022crystal} for an example application of the constitutive framework for modeling thermo-mechanical deformation in single crystal Ni-based superalloys.

\section{Code Implementation}
$\rho$-CP is developed as an application which utilizes and interfaces with the open source finite element framework, Multiphysics Object-Oriented Simulation Environment (MOOSE) \cite{permann2020moose}, for performing finite element simulations. MOOSE offers the ability to solve partial differential equations for multi-physics problems in massively parallel computing environments, using several thousands of processors \cite{permann2020moose}. Moreover, MOOSE has an already existing ecosystem for solving finite deformation mechanics problems using a Plug-n-Play system in the \texttt{TensorMechanics} module \cite{moose} and applying necessary boundary conditions, as well as interfacing with other physics environments, such as heat transfer, phase field, etc. In this regard, $\rho$-CP needs to be compiled alongside MOOSE to have access to the existing MOOSE libraries.

The crystal plasticity model in $\rho$-CP is implemented as an inherited class of the \texttt{ComputeStressBase} class from the above mentioned \texttt{TensorMechanics} module, which supplies an increment of the finite deformation gradient, $\bm F$, and the time step increment, $\Delta t$, at the Gauss points of a finite element mesh. The crystal plasticity model solves for the increment of stress and the tangent stiffness tensor corresponding to $\bm F$ due to the anisotropic elastic-plastic deformation, which are then passed back to MOOSE for global convergence computations. 

\begin{figure}[!htbp]
    \centering
	\includegraphics[scale=0.5]{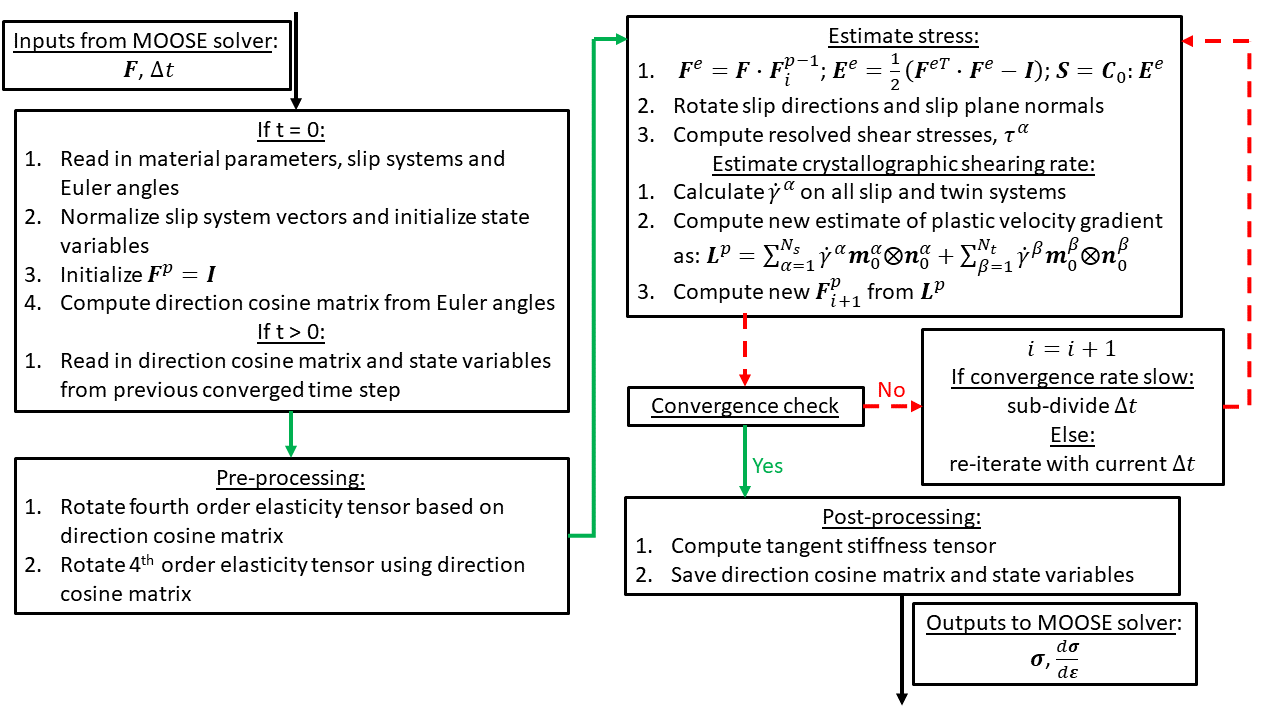}
	\caption{Algorithmic schematic of crystal plasticity solver.}
	\label{fig:algo_CP}
\end{figure}

The algorithmic steps involved in this are shown schematically in Figure \ref{fig:algo_CP}. Following the initialization, the crystal plasticity solver computes the initial guess of Cauchy stress, $\bm \sigma$, using elasticity calculations (cf. Section \ref{sec:elasticity}) and assuming $\bm F^{p}$ as the converged value of the corresponding tensor from the previous time step. The resolved shear stress, $\tau^{\alpha}$, is then computed on all slip and twin systems. If $|\tau^{\alpha} - \chi^{\alpha}|$ exceeds $s^{\alpha}_{a}$, then the corresponding crystallographic shearing rates are computed (similarly for twin systems). This process is repeated iteratively until a converged value of $\dot \gamma^{\alpha}$ is obtained on all slip and twin systems. Further, time step sub-incrementation is used when the rate of convergence is slow \cite{mcginty2001multiscale}. Based on these converged values, the stress and the tangent stiffness tensor are passed back to the FE solver for global convergence calculations. Also note that while the numerical implementation for the dislocation mean free path, $\lambda^{\alpha}$, has been performed with consideration for dislocations from all slip systems, the examples presented in Section \ref{results} assume that this term has contributions only from the primary slip system, as a first order approximation. This can be enabled or disabled by a parameter in the code.

Two separate classes, \texttt{DDCPStressUpdate} and \texttt{DDCPHCPStressUpdate}, have implemented these constitutive equations numerically in the $\rho$-CP repository. The former has implemented only the constitutive equations related to dislocation slip and may be used for cubic crystals, with identical material parameters on all slip systems, in the absence of twinning. In the later Sections, this \texttt{DDCPStressUpdate} class has been used to predict temperature- and strain rate-dependent deformation of fcc copper and bcc tantalum. Features such as on-the-run slip system assignment, and material properties and model parameter assignment allow reuse of the same class for materials with different crystal structures. The \texttt{DDCPHCPStressUpdate} class implements constitutive equations related to both dislocation slip and twinning. Further, it allows assignment of different material properties to different slip and twin systems, such as prismatic, basal, pyramidal, etc. An example application has been demonstrated for hcp magnesium.

\begin{figure}[!htbp]
    \centering
	\includegraphics[scale=0.5]{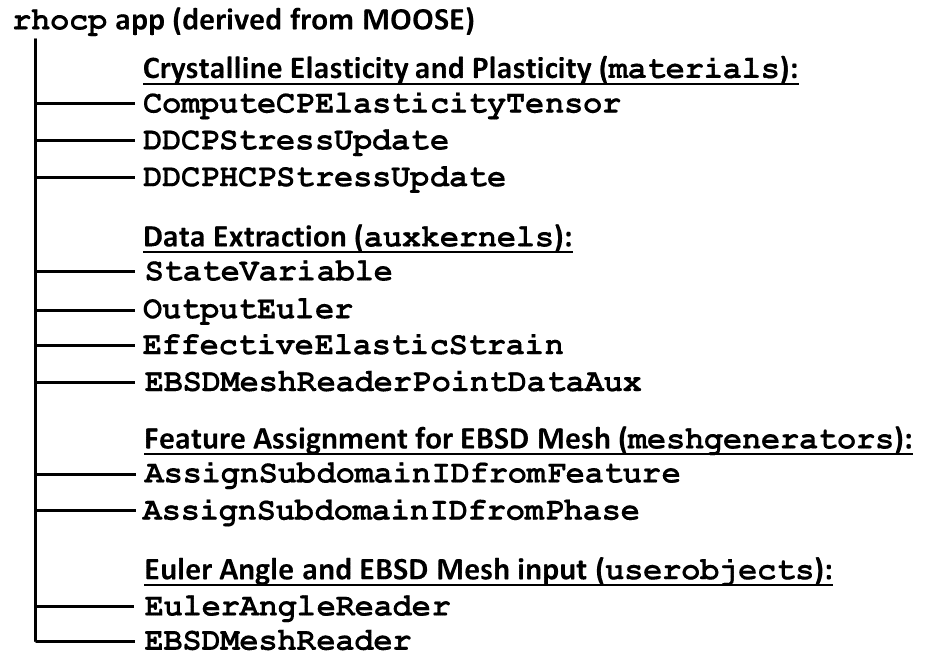}
	\caption{Class structure of $\rho$-CP application.}
	\label{fig:class_structure}
\end{figure}

The overall class structure of the $\rho$-CP application is shown in Figure \ref{fig:class_structure}. While there is a vast library of classes already existing in the MOOSE repository, these additional classes have been implemented for ease of data transfer between the MOOSE and $\rho$-CP classes, as well as pre-processing and post-processing of information from the crystal plasticity solver. $\rho$-CP also allows the user to utilize the restart features in MOOSE by saving all history-dependent variables required by the CP solver as state variables. This prevents the need for starting from scratch those simulations that were unintentionally terminated due to hardware or software related issues beyond the user's control. The restart feature is also useful in cases where the finite element solver does not converge and modification of the simulation convergence parameters is needed mid-way through the simulations.

\section{Example Applications}
\label{results}
In this Section, we demonstrate example applications of $\rho$-CP to simulate the deformation behavior of magnesium, copper and tantalum over a range of loading conditions. While the first two examples are for room temperature deformation, the tantalum simulations are performed over a range of deformation temperatures and strain rates to demonstrate the constitutive model's ability to predict such effects.

\subsection{Magnesium}
Magnesium, with low density and high specific strength, is a structural material of interest for automotive and other light-weighting applications \cite{shi2022anisotropy}. However, anisotropic mechanical properties and limited ductility are known issues associated with this material \cite{agnew2001application}. Magnesium has a hexagonal close packed crystal structure. Depending on the crystal orientation and loading conditions, different deformation modes may be active in magnesium. In the present study, we have considered the following allowable deformation modes for room temperature deformation: 3 basal slip systems ($(0001)<11 \bar 20>$), 3 prismatic slip systems ($\{10 \bar 10\}<11 \bar 20>$), 6 pyramidal <a> slip systems ($\{10 \bar 11\}<11 \bar 20>$), 6 pyramidal <c+a> slip systems ($\{11 \bar 22\}<11 \bar 23>$), and 6 tensile twinning systems ($\{10 \bar 12\}<\bar 1011>$) \cite{zhang2012phenomenological}.

\begin{figure}[!htbp]
    \centering
	\includegraphics[scale=0.5]{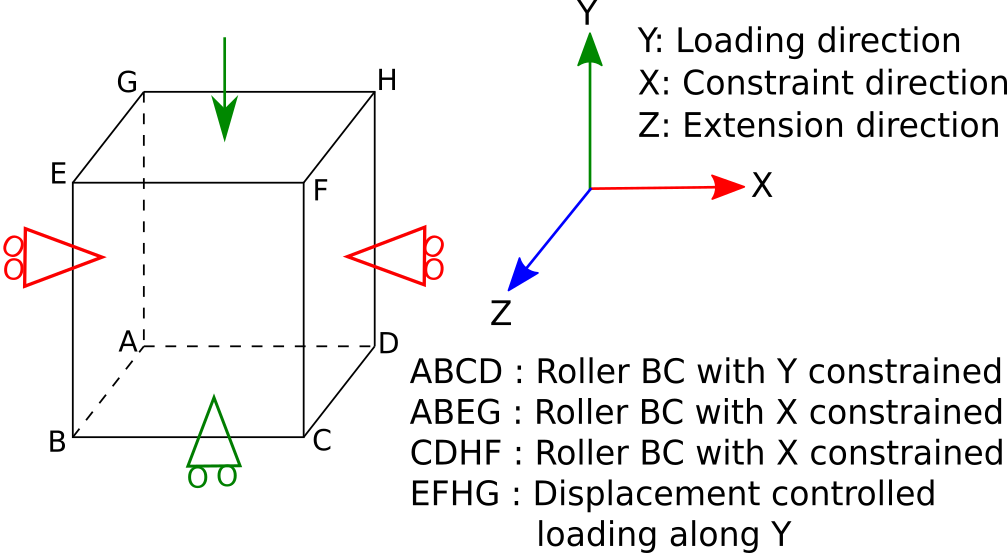}
	\caption{Schematic of loading and boundary conditions used for plane strain compression of magnesium single and polycrystals.}
	\label{fig:mg_sx_psc}
\end{figure}

Channel die compression experiments, representative of plane strain compression deformation, have been previously performed on single crystal and polycrystalline magnesium to characterize the deformation anisotropy and texture-dependent response \cite{kelley1968plane, kelley1968deformation}. We have simulated deformation under representative loading and boundary conditions to replicate these experiments. For the single crystal simulations, a cube-shaped domain having 2 hexahedral finite elements per side (total 8 elements) was considered. Note that all the simulation results presented in this and the following Sections have used finite elements with linear interpolation. The bottom face was constrained to move along the y-direction, while displacement-controlled compressive loading was applied on the top face at a nominal strain rate of $1 \times 10^{-3}$ /s. Further, motion along the x-direction was constrained on the lateral faces to simulate the die constraint. Note that we have not considered any frictional effects between the deformation specimens and the channel die. The sample is free to flow along the z-direction. These loading and boundary conditions are schematically shown in Figure \ref{fig:mg_sx_psc}. 

\begin{figure}[!htbp]
    \centering
	\includegraphics[scale=0.65]{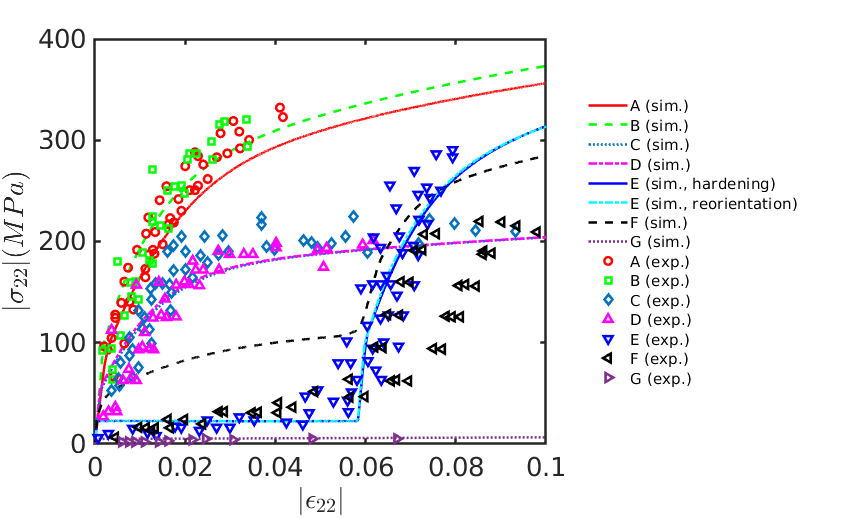}
	\caption{Model predictions and comparison with the corresponding experimental stress-strain response along the loading direction for plane strain compression of magnesium single crystals loaded in seven different orientations. Solid lines represent model predictions, while the experimental data points are represented using open symbols. The experimental data were obtained from \cite{kelley1968plane, graff2007yielding}.}
	\label{fig:mg_sx_stress_strain}
\end{figure}

\begin{figure}[!htbp]
    \centering
	\includegraphics[scale=0.47]{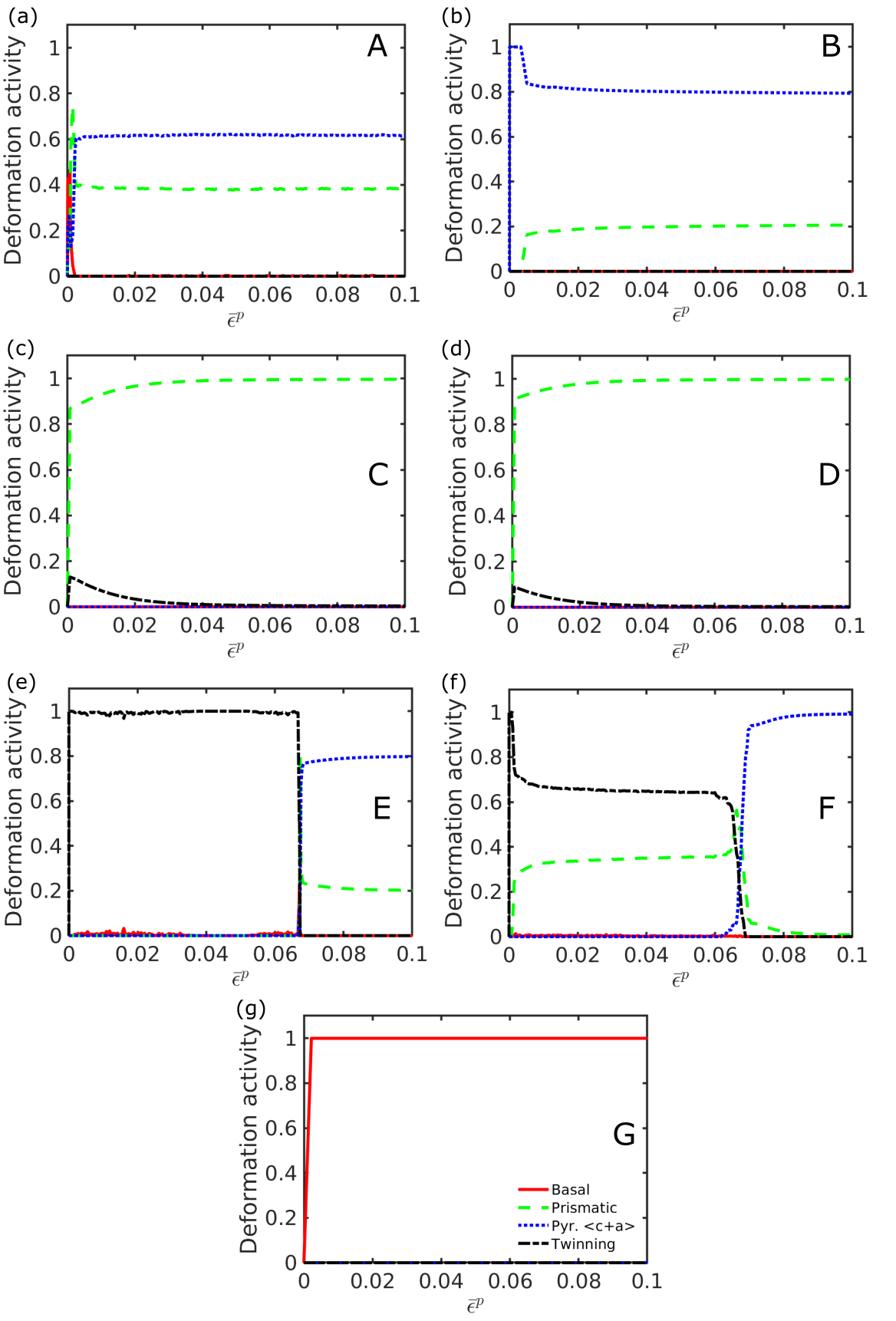}
	\caption{Model predictions of deformation mode activities for hcp magnesium single crystals loaded along orientations (a) A, (b) B, (c) C, (d) D, (e) E, (f) F, and (g) G, respectively, corresponding to the stress-strain responses shown in Figure \ref{fig:mg_sx_stress_strain}.}
	\label{fig:mg_sx_slip_activity}
\end{figure}

Simulations have been performed for seven distinct crystal orientations, which are expected to have one primary deformation mode active for each case. This allows us to individually calibrate the single crystal constitutive model parameters for each deformation mode. The Euler angles (in Bunge notation) for these seven orientations are given in Table \ref{tab: mg_sx_orientations}. The constitutive model was first calibrated to predict the orientation with basal slip, followed by prismatic and pyramidal <c+a> slip, respectively. Finally, model parameters related to tensile twinning were calibrated. The anisotropic elastic constants for magnesium are given in Table \ref{Elastic_constants}, while the constitutive model parameters related to room temperature plastic deformation are given in Table \ref{tab: mg_params}. Note that we have not used the model to predict cyclic deformation of magnesium. Accordingly, the slip system-level backstress has been assumed to be absent for this example. Figure \ref{fig:mg_sx_stress_strain} shows the model predictions of the stress-strain response as compared with the corresponding experimental data \cite{kelley1968plane, graff2007yielding}. The deformation mode activity for these loading orientations is also shown in Figure \ref{fig:mg_sx_slip_activity}. Also note that for orientation E, we have plotted the predicted stress-strain response using both the hardening model (cf. Equation \ref{twin_hardening}) and the twin reorientation model (cf. Equation \ref{twin_reorient}).

\begin{table}[!htbp]
  \begin{center}
    \caption{Euler angles (in Bunge notation) for the magnesium single crystal simulations.}
    \label{tab: mg_sx_orientations}
    \begin{tabular}{l c c c}
     \hline
       Orientation & $\phi_{1}$ ($^{\circ}$) & $\Phi$ ($^{\circ}$) & $\phi_{2}$ ($^{\circ}$) \\ 
       \hline
       A & 0 & 90 & 30 \\
       B & 0 & 90 & 0 \\
       C & 90 & 90 & 90 \\
       D & 60 & 90 & 90 \\
       E & 0 & 0 & 0\\
       F & 30 & 0 & 0 \\
       G & 30 & 45 & 0 \\
       \hline
     \end{tabular}
  \end{center}
\end{table}

As can be seen, the model predictions compare reasonably well with the experimental counterparts for all orientations, except orientation F (this is discussed later). Orientations A and B show dominant pyramidal <c+a> slip activity, with some prismatic slip activity. Orientations C and D show primarily prismatic slip activity. Orientations E and F show dominant tensile twinning activity prior to the point of twin reorientation, while it is dominated by prismatic slip subsequently. Finally, orientation G shows primarily basal slip activity. Note that pyramidal <a> slip was not observed in either case, and has not been plotted here. As mentioned earlier, predictions from both the twin hardening and the twin reorientation models are shown Figure \ref{fig:mg_sx_stress_strain}. While both these models give reasonable comparison with the experimental data, the twin reorientation model was found to give convergence issues due to the sudden change in crystal orientation and associated elastic stiffness during reorientation. Accordingly, only the twin hardening model was used for the polycrystal simulations presented next. As for orientation F, it was found that some prismatic slip activity occurs along with tensile twinning. In our rather simplified constitutive models for twinning and slip, we have not considered any interactions between the twinning and slip systems. This could have contributed to the higher predicted flow stress as compared to the experimental data for orientation F in the twinning regime. We have also verified that increasing the number of elements in the simulation domain makes the macroscopic response marginally compliant and the predicted stress at a given strain is lower by only $< 2\%$ using a larger number of elements, up to 125 elements (results not presented here). However, this is still not able to predict the experimentally observed response for orientation F.

\begin{table}[!htbp]
   \begin{center}
     \caption{Anisotropic elastic constants for hcp magnesium \cite{zhang2012phenomenological}, fcc OFHC copper \cite{simmons1971single, wu1996simulation} and bcc tantalum \cite{kothari1998elasto}. Note that temperature effects on the elastic constants are only considered for tantalum, while the room temperature values are presented for magnesium and OFHC copper.}
     \begin{tabular}{c c c c}
      \hline
         Parameter & Magnesium & Copper & Tantalum \\
         \hline
         $C_{11}\left(GPa\right)$ & 59.4  & 170.0  & 268.2 \\ 
         $\frac{dC_{11}}{dT}\left(\frac{GPa}{K}\right)$ & - & -  &  0.024\\ 
         $C_{12}\left(GPa\right)$ & 25.6 & 124.0  & 159.6\\ 
         $\frac{dC_{12}}{dT}\left(\frac{GPa}{K}\right)$ & - & -  &  0.011\\ 
         $C_{13}\left(GPa\right)$ & 21.4 & -  & -\\ 
         $\frac{dC_{13}}{dT}\left(\frac{GPa}{K}\right)$ & - & -  &  -\\ 
         $C_{33}\left(GPa\right)$ & 61.6 & -  & - \\ 
         $\frac{dC_{33}}{dT}\left(\frac{GPa}{K}\right)$ & - & -  &  - \\ 
         $C_{44}\left(GPa\right)$ & 16.4 & 75.0  & 87.1\\ 
         $\frac{dC_{44}}{dT}\left(\frac{GPa}{K}\right)$ & - & -  & 0.015 \\ 
         $G\left(GPa\right)$ & 16.4 & 41.5  & 87.1 \\ 
         $\frac{dG}{dT}\left(\frac{GPa}{K}\right)$ & - & - & 0.015\\
         \hline
         \end{tabular}
     \label{Elastic_constants}
   \end{center}
\end{table}

\begin{table}[!htbp]
  \begin{center}
    \caption{Constitutive model parameters for magnesium.}
    \label{tab: mg_params}
    \begin{tabular}{l c c c c c}
     \hline
      
       Parameter & Basal & Prismatic & Pyramidal $\langle a \rangle$ & Pyramidal $\langle c+a \rangle$ & Tensile twin \\ 
       \hline
       $b^{\alpha} \left(nm\right)$ & 0.321 & 0.321 & 0.612 & 0.612 & - \\ 
       
       $\dot \gamma^{\alpha}_{0s} \left(s^{-1}\right)$ & $1.0$  & $1.0 \times 10^{-2}$  & $40$ & $1.0 \times 10^{-3}$ & -\\

       $\Delta F^{\alpha}$ & 0.11 $Gb^{3}$ & 0.2 $Gb^{3}$ & 1.43 $Gb^{3}$ & 0.1 $Gb^{3}$ & - \\

       $p^{\alpha}$ & 0.2 & 0.2 & 0.3 & 0.2 & - \\
       
       $q^{\alpha}$ & 1.7 & 1.7 & 1.5 & 1.7 & - \\
       
       $\tau^{\alpha}_{0s} \left(\mathrm{MPa}\right)$ & 2.0 & 21.0 & 50.0 & 38.0 & - \\

       $s^{\alpha}_{t} \left(\mathrm{MPa}\right)$ & 5.0 & 21.0  & 100.0  & 5.0 & - \\

       $k^{\alpha}_{\rho}$ & 0.35 & 0.535 & 0.35 & 0.35 & - \\
       
       $A^{\alpha\alpha}$, $A^{\alpha\zeta}$ & 1.0, 0.2 & 1.0, 0.2 & 1.0, 0.2 & 1.0, 0.2 & - \\
       
       $\rho^{0}_m \left(m^{-2}\right)$ & $1 \times 10^{10}$  &$1 \times 10^{10}$ & $1 \times 10^{10}$ & $1 \times 10^{10}$ & - \\
       
       $\rho^{0}_i \left(m^{-2}\right)$ & $1 \times 10^{10}$ & $1 \times 10^{10}$ & $1 \times 10^{10}$ & $1 \times 10^{10}$  & - \\
       
       $k^{\alpha}_{M}$ & 0.0017 & 1.0 & 1.0 & 3.5 & - \\
       
       $R^{\alpha}_c \left(nm\right)$ & 19.386 & 19.386 & 19.386 & 36.462 & - \\
       
       $k^{\alpha}_{I}$ & 0.0015 & 0.98 & 0.8 & 3.4 & - \\
       
       $k^{\alpha}_{D}$ & 0.5 & 180 & 500 & 350 & - \\
       
       $\dot \gamma^{\alpha}_{0t} \left(s^{-1}\right)$ & - & - & - & - & $1.0 \times 10^{-3}$ \\ 

       $\tau^{\alpha}_{0t} \left(\mathrm{MPa}\right)$ & - & - & - & - & 2.0 \\

       $D^{\alpha}_{0} \left(\mathrm{MPa}\right)$ & - & - & - & - & 10.0 \\
       
       $m^{\alpha}$ & - & - & - & - & 20 \\
       
       $\gamma^{\alpha}_{tw}$ & - & - & - & - & 0.1289 \\

       $h^{\alpha}_{0} \left(\mathrm{MPa}\right)$ & - & - & - & - & 1000 \\
       
       $m^{\alpha}_{th}$ & - & - & - & - & 10 \\
       \hline
    \end{tabular}
  \end{center}
\end{table}

\begin{figure}[!htbp]
    \centering
	\includegraphics[scale=0.33]{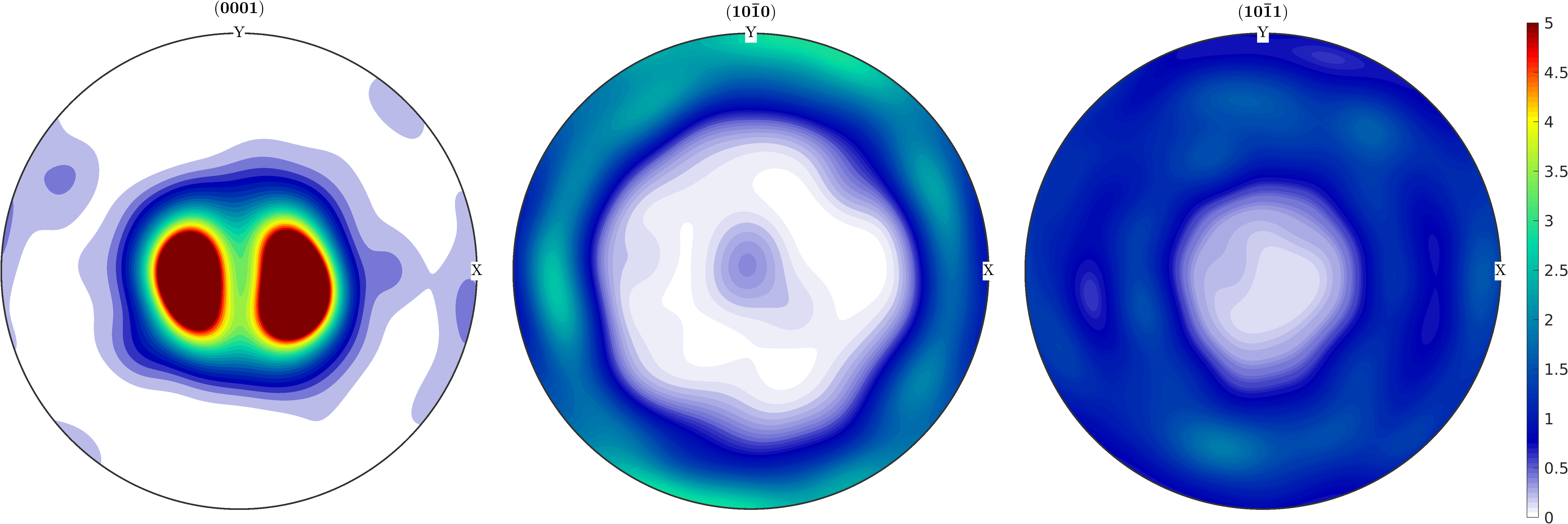}
	\caption{Initial texture, with 512 orientations, used for the magnesium polycrystal simulations, plotted in terms of the (0001), $(10 \bar 1 0)$, and $(10 \bar 1 1)$ pole figures. Directions X, Y and Z correspond to the rolling, transverse and normal directions of the rolled sheet.}
	\label{fig:mg_rolled_texture}
\end{figure}

\begin{figure}[!htbp]
    \centering
	\includegraphics[scale=0.55]{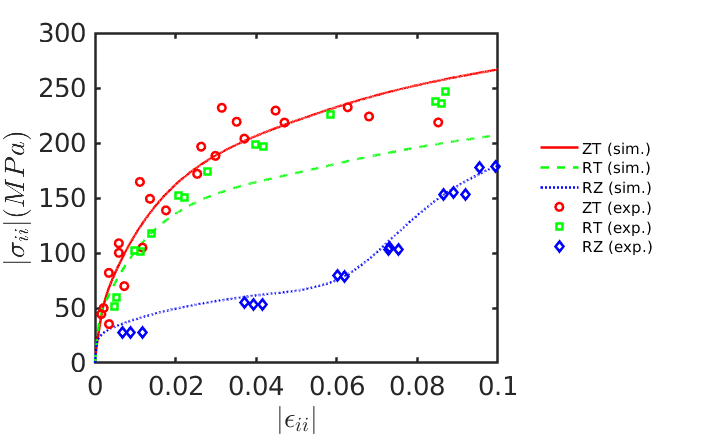}
	\caption{Model predictions and comparison with the corresponding experimental stress-strain response along the loading direction for plane strain compression of polycrystalline hcp magnesium. Solid lines represent model predictions, while the experimental data points are represented using open symbols. The experimental data were obtained from \cite{kelley1968deformation, graff2007yielding}.}
	\label{fig:mg_px_stress_strain}
\end{figure}

\begin{figure}[!htbp]
    \centering
	\includegraphics[scale=0.55]{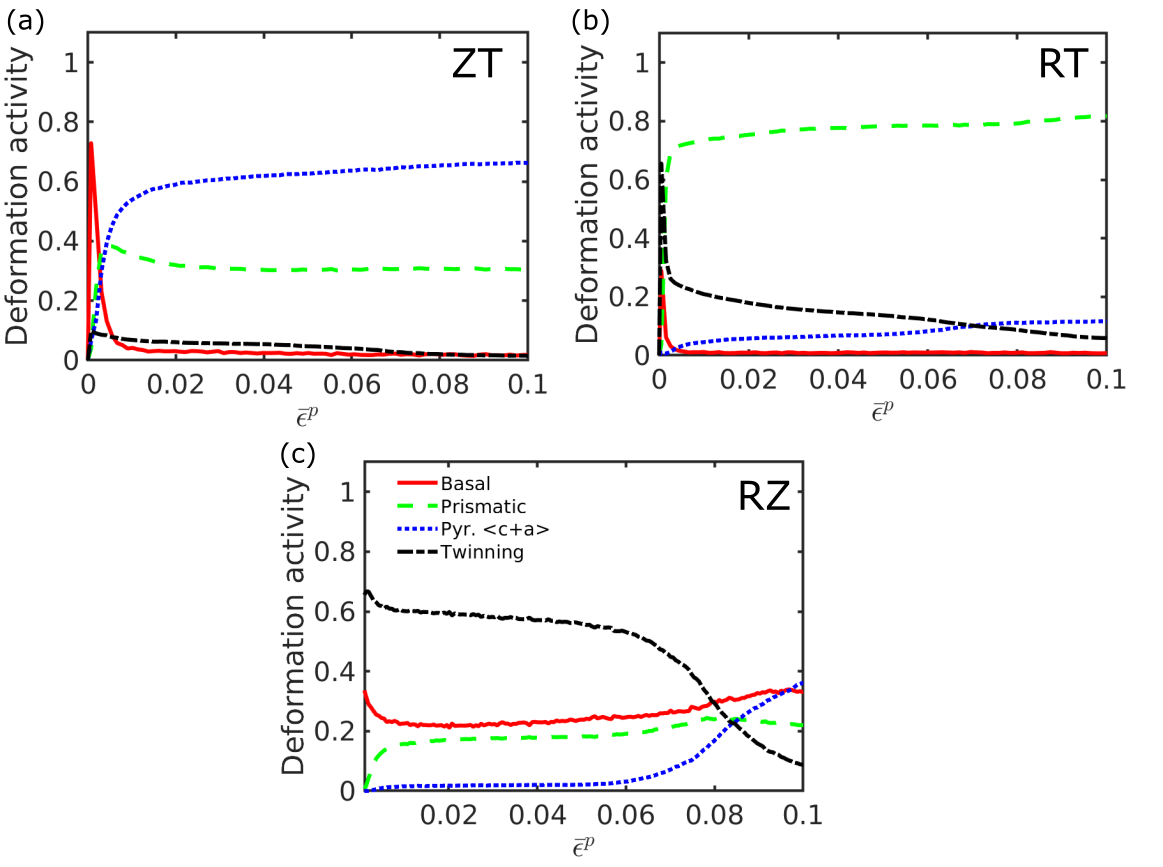}
	\caption{Model predictions of deformation mode activities for magnesium polycrystals loaded along orientations (a) ZT, (b) RT, and (c) RZ, respectively, corresponding to the stress-strain responses shown in Figure \ref{fig:mg_px_stress_strain}.}
	\label{fig:mg_px_slip_activity}
\end{figure}

The calibrated model was used to predict the orientation-dependent response of textured polycrystalline magnesium. For these simulations, a representative texture comprised of 512 orientations was first created synthetically such that it qualitatively resembles that of a rolled magnesium plate \cite{agnew2001application}. This initial texture is shown in Figure \ref{fig:mg_rolled_texture} in terms of the (0001), $(10 \bar 1 0)$, and $(10 \bar 1 1)$ pole figures. Essentially, there is a strong concentration of c-axis poles along the normal (z) direction. 3D finite element simulations were performed using a 512 element mesh, such that each element was assumed to represent one grain orientation. Boundary conditions similar to that in Figure \ref{fig:mg_sx_psc} were then used to simulate channel die compression of the rolled magnesium plate loaded along different directions. The only difference in these simulations is that the loading and constraint directions were changed based on the texture, rather than rotating the crystal orientations/texture (as done for the magnesium single crystals). Model predictions, as compared with the experimental counterparts \cite{kelley1968deformation, graff2007yielding}, are shown for three different loading orientations in Figure \ref{fig:mg_px_stress_strain}. The corresponding deformation mode activities are shown in Figure \ref{fig:mg_px_slip_activity}. In these simulations, the direction R represents the rolling (x) direction, T represents the transverse (y) direction, and z represents the normal direction with respect to the rolled plate. Further, the nomenclature ZT indicates that the polycrystal is loaded along the Z direction, while it is constrained along the T direction. In this case, the polycrystal is free to expand along the third direction, R. Loading and constraint directions for the other two loading orientations, RT and RZ, may be interpreted similarly.  

It can be seen from Figure \ref{fig:mg_px_stress_strain} that while the model predicts the response for ZT and RT cases with reasonable accuracy, a relatively lower flow stress is predicted for the RT case as compared to the experimental data. The deformation activity plots in Figure \ref{fig:mg_px_slip_activity} show that pyramidal <c+a> slip is dominant for the ZT case, prismatic slip is dominant for the RT case, while tensile twinning is dominant for the RZ case. It should also be noted that unlike the single crystal deformation mode activity plots (cf. Figure \ref{fig:mg_sx_slip_activity}), secondary deformation modes are also present for the magnesium polycrystals. For example, some tensile twinning is observed for the RT case, in addition to prismatic slip. As discussed earlier (cf. Section \ref{sec:constitutive}), we have not considered hardening on the twinning systems (except due to twin reorientation), neither have we considered twin-slip interactions \cite{zhang2012phenomenological}. Neglect of these hardening mechanisms may have contributed to the under-prediction of the flow stress for the RT case. Additional mechanisms may be incorporated in future work to predict more accurately the effect of hardening due to twin-slip interactions. Nonetheless, we have demonstrated the ability of the model to predict the orientation-dependent single crystal response of magnesium and then used the same constitutive model parameters to predict the texture-dependent polycrystalline response.

\subsection{Copper}

We next use the model to predict the mechanical response of polycrystalline Oxygen-Free High Conductivity (OFHC) copper. In this example, the primary objective is to demonstrate the model's ability to predict strain rate-dependent deformation as well as backstress evolution under cyclic loading at room temperature. OFHC copper has a face-centered cubic crystal structure and 12 possible octahedral slip systems ($\{ 111 \}<110>$). Further, it was assumed that twinning systems are not active during room temperature deformation of copper.

\begin{figure}[!htbp]
    \centering
	\includegraphics[scale=0.5]{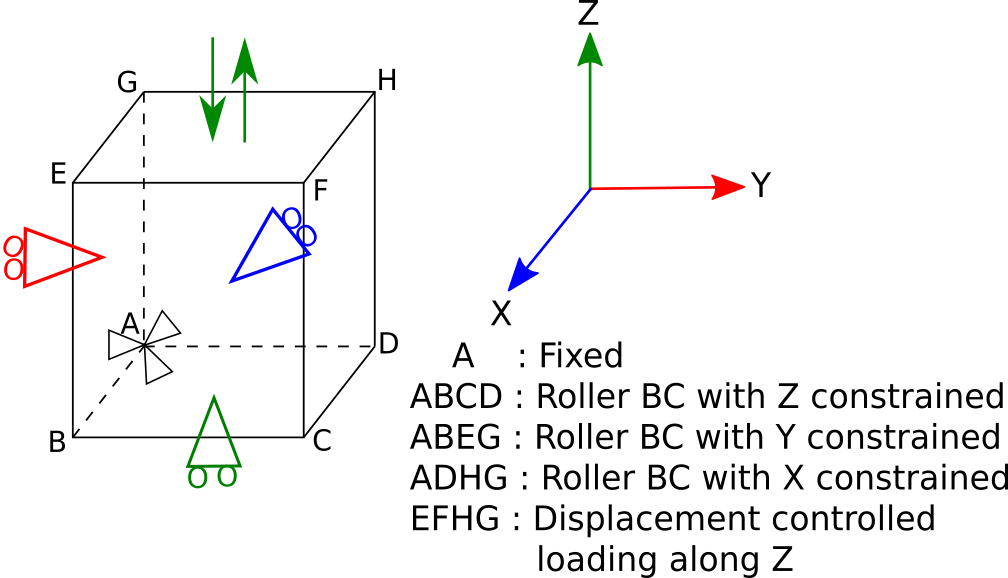}
	\caption{Schematic of loading and symmetric boundary conditions used for the OFHC copper simulations.}
	\label{fig:symmetricBC}
\end{figure}

For these simulations a pseudo-random texture, comprised of 64 orientations, was used. A cube-shaped simulation domain, with 8 3D hexahedral elements per grain (total 512 elements), was used for these simulations. Symmetric boundary conditions were used for these simulations. Displacements normal to each of the back faces of the cubic domain were constrained and the corner node common to these three faces was constrained in all degrees of freedom. Displacement-controlled loading was applied on the front face along the z-direction. This is schematically shown in Figure \ref{fig:symmetricBC}.

\begin{figure}[!htbp]
    \centering
	\includegraphics[scale=0.55]{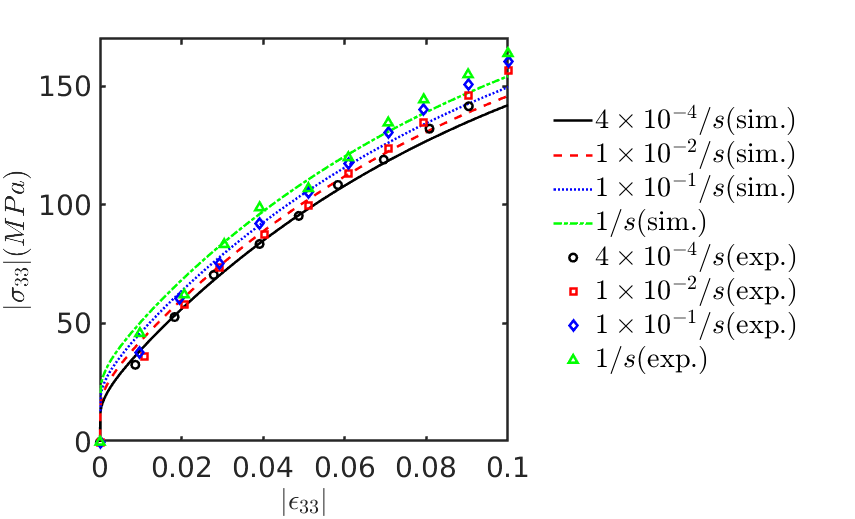}
	\caption{Model predictions and comparison with the corresponding experimental response for uniaxial compression tests at different strain rates for polycrystalline fcc OFHC copper at 298 K. Solid lines represent model predictions, while the experimental data points are represented using open symbols. The experimental data were taken from \cite{tanner1999deformation}.}
	\label{fig:cu_strain_rate}
\end{figure}

The constitutive model was first fitted to predict the strain rate-dependent mechanical response under uniaxial compression. These results are shown in Figure \ref{fig:cu_strain_rate}, along with the comparison to the experimental data \cite{tanner1999deformation}. The anisotropic elastic constants for copper are given in Table \ref{Elastic_constants}, while the constitutive model parameters for room temperature plastic deformation are given in Table \ref{param_table}. It can be seen that the strain rate effect, over a range of four orders of magnitude, while weak, is reasonably predicted by the model up to 0.1 applied strain. The present form of the model does not account for strain rate (or temperature) effects on the hardening response, and primarily the initial yield stress is affected. This could be addressed in future work by allowing the $s^{\alpha}_t$ term and the dislocation evolution parameters to evolve during deformation as well (cf. Equation \ref{eqn:flow rule slip}).

\begin{figure}[!htbp]
    \centering
	\includegraphics[scale=0.55]{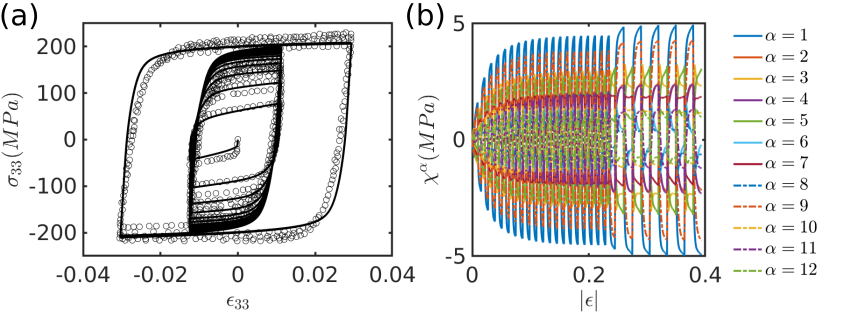}
	\caption{(a) Model predictions and comparison with the corresponding experimental response for cyclic loading for polycrystalline fcc OFHC copper at 298 K. Solid lines represent model predictions, while the experimental data points are represented using open circles. The experimental data were taken from \cite{tanner1998thesis}. (b) Evolution of the element-averaged slip system-level backstress during the cyclic loading simulations. The x-axis represents the nominal "axial" strain additively summed over the loading cycles.}
	\label{fig:cu_cyclic}
\end{figure}

In order to highlight the model's ability to predict backstress-associated hardening, we have simulated cyclic loading according to the experiments given in \cite{tanner1998thesis}. Similar boundary conditions as for uniaxial compression were used, while the displacement rate on the loading face was adjusted appropriately to simulate tensile and compressive loading. Fully reversed compression-tension was first simulated for 20 cycles up to a strain of $\pm 0.01$, followed by 5 cycles up to a strain of $\pm 0.03$ at a nominal strain rate of $1 \times 10^{-4}$ /s. The model prediction and its comparison with experimental data is shown in Figure \ref{fig:cu_cyclic} (a). The corresponding evolution of the slip system-level backstress with nominal strain for the different slip systems is shown in Figure \ref{fig:cu_cyclic} (b). 

It can be seen that the model predicts the cyclic response with reasonable accuracy. While there is significant hardening during the first 5-6 cycles, the hardening tends to saturate during subsequent cycles and even at the higher strain amplitude. This is evident both from the cyclic stress-strain response as well as from the backstress evolution plots. Further, the saturated value of the slip system-level backstress is less than 5 MPa even after loading at the high strain amplitude of 0.03. Given that the Taylor factor is expected to be of the order of 3, the overall contribution of the slip system-level backstress to the macroscopic flow stress is expected to be less than 15 MPa ($\leq 10 \%$). Thus, isotropic hardening due to the mobile and immobile dislocation densities is the dominant contributor to the flow stress of the polycrystalline copper under consideration, rather than slip system-level backstress. Subsequent to the first 20 cycles, there is some deviation from the experimental response during the elastic loading / unloading part for the higher strain amplitude. However, note that the peak stresses at the end of the loading cycle are still comparable between simulations and experiments.  

\begin{table}[!htbp]
   \begin{center}
     \caption{Constitutive model parameters for OFHC copper and tantalum. Note that the tantalum model was not calibrated to any cyclic loading experiments and it was assumed that slip system-level backstress is not present.}
     \begin{tabular}{c c c}
      \hline
         Parameter & Copper & Tantalum \\
         \hline
         $b^{\alpha} \left(nm\right)$ & 0.256 & 0.286 \\ 
         $\dot{\gamma}_{0s}^\alpha \left(s^{-1}\right)$ & $4\times10^{6}$ & $1.73\times10^{7}$ \\
         $\Delta F^{\alpha}$ & 0.25$Gb^{3}$ & 0.2$Gb^{3}$ \\
         $p^{\alpha}$ & 0.35 & 0.28 \\ 
         $q^{\alpha}$ & 1.3 & 1.38 \\ 
         $\tau^{\alpha}_{0s} \left(\mathrm{MPa}\right)$ & 0.0  & 24 \\ 
         $s^{\alpha}_{t}  \left(\mathrm{MPa}\right)$ & 38  & 386 \\
         $k_{\rho}^{\alpha}$ & 0.2 & 0.3\\
         $A^{\alpha\alpha}$, $A^{\alpha\zeta}$ & 1, 0.1 & 1, 0.1 \\
         $\rho^{0}_{m}\left(m^{-2}\right)$  & $5\times10^{11}$ & $1\times10^{11}$ \\
         $\rho^{0}_{i}\left(m^{-2}\right)$ &$5\times10^{11}$ & $1\times10^{11}$\\
         $k_{M}^{\alpha}$ & 0.13 & 0.05 \\
         $R^{\alpha}_c\left(nm\right)$ & 1.53& 1.488 \\
         $k_{I}^{\alpha}$ & 0.12 & 0.045\\
         $k_{D}^{\alpha}$ & 40 & 20 \\ 
         $k^{\alpha}_{\chi 1}$ & 1100 & - \\
         $k^{\alpha}_{\chi 2}$ & 1000 & - \\
         \hline
         \end{tabular}
     \label{param_table}
   \end{center}
\end{table}

\subsection{Tantalum}

The remaining examples in this manuscript are demonstrated for tantalum. Tantalum is refractory metal with superior high temperature mechanical properties and finds use in structural applications \cite{cardonne1995tantalum}. It has a body-centered cubic crystal structure. Based on prior studies \cite{kothari1998elasto, lee2023deformation}, we have assumed that 12 $\{ 110 \} <111>$ and 12 $\{ 112 \} <111>$ are the available slip systems in tantalum.

\subsubsection{Temperature and Strain Rate Effects}
A pseudo-random texture comprised of 64 orientations was first used to predict the temperature and strain rate effects on the yield stress. Similar to the copper simulations, a cube-shaped simulation domain with 8 3D hexahedral elements per grain (total 512 elements) and symmetric boundary conditions was used.

\begin{figure}[!htbp]
    \centering
	\includegraphics[scale=0.55]{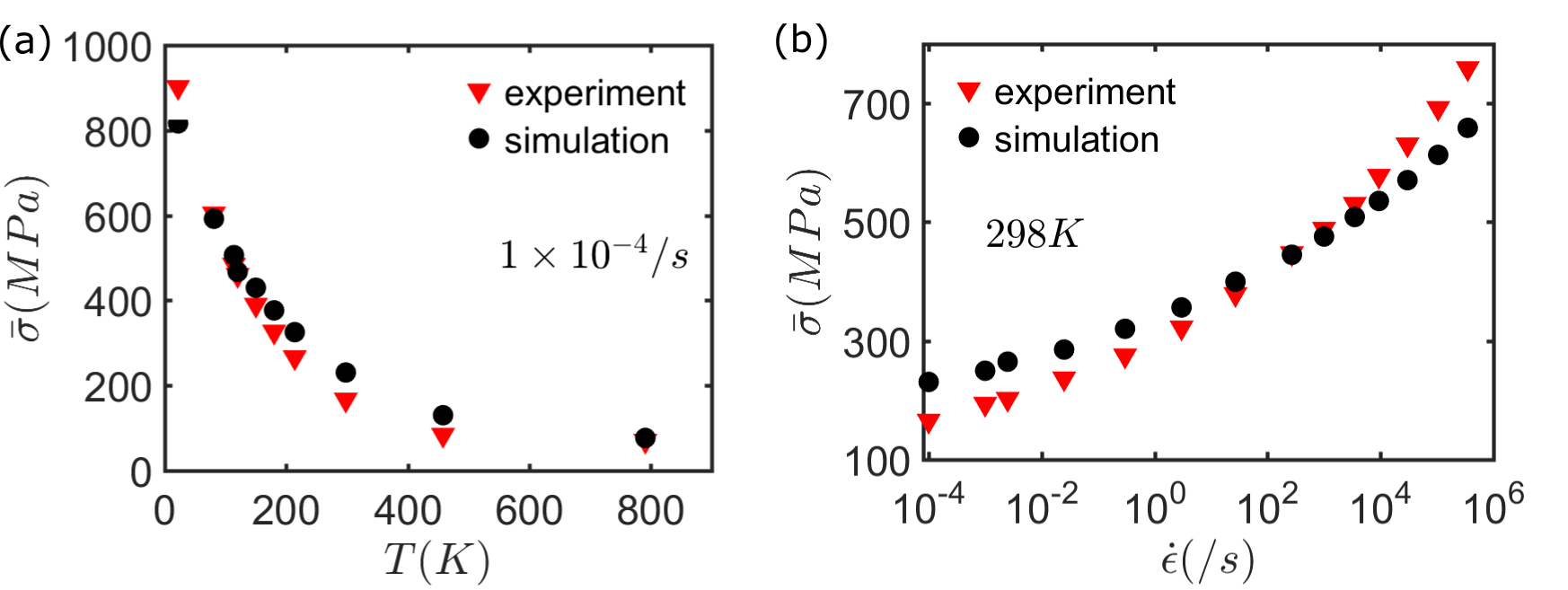}
	\caption{Comparison of the predicted yield stress ($0.2 \%$ offset) with the experimental counterparts as a function of (a) temperature, and (b) strain rate during tensile deformation of polycrystalline bcc tantalum. The experimental data were taken from \cite{hoge1977temperature}.}
	\label{fig:ta_ys}
\end{figure}

The flow parameters, $\dot \gamma^{\alpha}_{0s}$, $\Delta F^{\alpha}$, $p^{\alpha}$ and $q^{\alpha}$, and the intrinsic lattice resistances, $\tau^{\alpha}_{0s}$ and $s^{\alpha}_{t}$, were first calibrated to predict the temperature- and strain rate-dependent yield stress. For simplicity, the same parameters were assumed for both $\{ 110 \} <111>$ and $\{ 112 \} <111>$ slip systems. Note that these parameters can be estimated with reasonable accuracy using analytical calculations, without the need for running computationally expensive CPFE simulations. Model predictions of the temperature-dependent yield stress at a quasi-static strain rate of $1 \times 10^{-4}$ /s are shown in Figure \ref{fig:ta_ys} (a), while those performed for different strain rates, spanning more than 10 orders of magnitude, at 298 K are shown in Figure \ref{fig:ta_ys} (b) and compared with the corresponding experimental data. The experimental data were taken from \cite{hoge1977temperature}. It can be seen that the model is able to predict the temperature-dependent yield stress across the entire range (22 K - 791 K) with reasonable accuracy, while there is some discrepancy in the prediction of the strain rate-dependent yield stress, especially at the extremes. As will be seen next, despite this discrepancy, the flow stress predictions appear reasonable. Before moving forward, it should be noted that we have not considered the effect of non-Schmid stresses on the yield behavior of tantalum. Such effects may be expected to influence to be dominant on the single crystal yield behavior \cite{cho2018anomalous}, while we have primarily focused on the deformation behavior of polycrystalline tantalum in this study.

\begin{figure}[!htbp]
    \centering
	\includegraphics[scale=0.55]{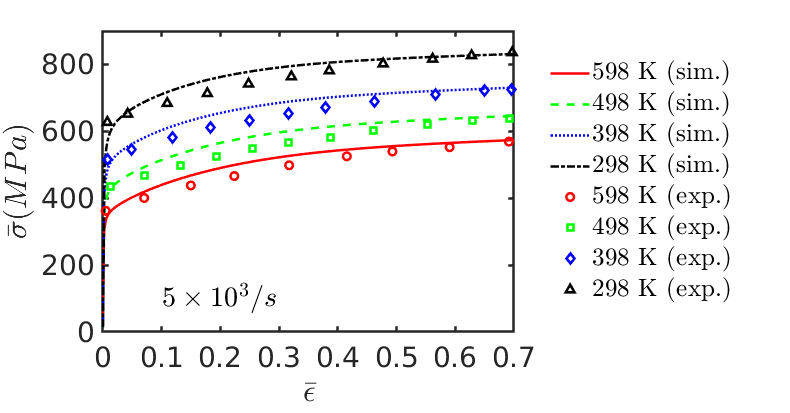}
	\caption{Comparison of predicted stress-strain response with the experimental counterparts during uniaxial compression for different temperatures at an applied strain rate of $5 \times 10^{3} /s$ for polycrystalline bcc tantalum. Solid lines represent model predictions, while the experimental data points are represented using open symbols. The experimental data were taken from \cite{nemat1997direct, kothari1998elasto}.}
	\label{fig:ta_stress_strain}
\end{figure}

The model was then used to predict the flow stress under uniaxial compression at an imposed nominal strain rate of $5 \times 10^{3}$ /s for four different temperatures. The hardening response is primarily influenced by the dislocation hardening coefficient, $k^{\alpha}_{\rho j}$, and the dislocation evolution parameters, $k^{\alpha}_{M}$, $R^{\alpha}_c$, $k^{\alpha}_{I}$ and $k^{\alpha}_{D}$ (again assumed to be the same for both  $\{ 110 \} <111>$ and $\{ 112 \} <111>$ slip systems.). There was some trial and error involved in estimating the dislocation evolution parameters. The fitted values of these parameters are given in Table \ref{param_table}. The comparison of the predicted flow stress with the corresponding experimental data from \cite{nemat1997direct, kothari1998elasto} up to 0.7 effective strain is shown in Figure \ref{fig:ta_stress_strain}. It can be seen that there is qualitative concurrence of the predicted flow stress over the entire range of deformation for all four temperatures. These results highlight the ability of the model to predict the yield and flow stress over a range of temperatures and strain rates for relatively large strains. In the following Sections, the constitutive model for tantalum is used to predict different microstructure and substructure evolution characteristics.

\subsubsection{Texture and Substructure Evolution}

\begin{figure}[!htbp]
    \centering
	\includegraphics[scale=0.45]{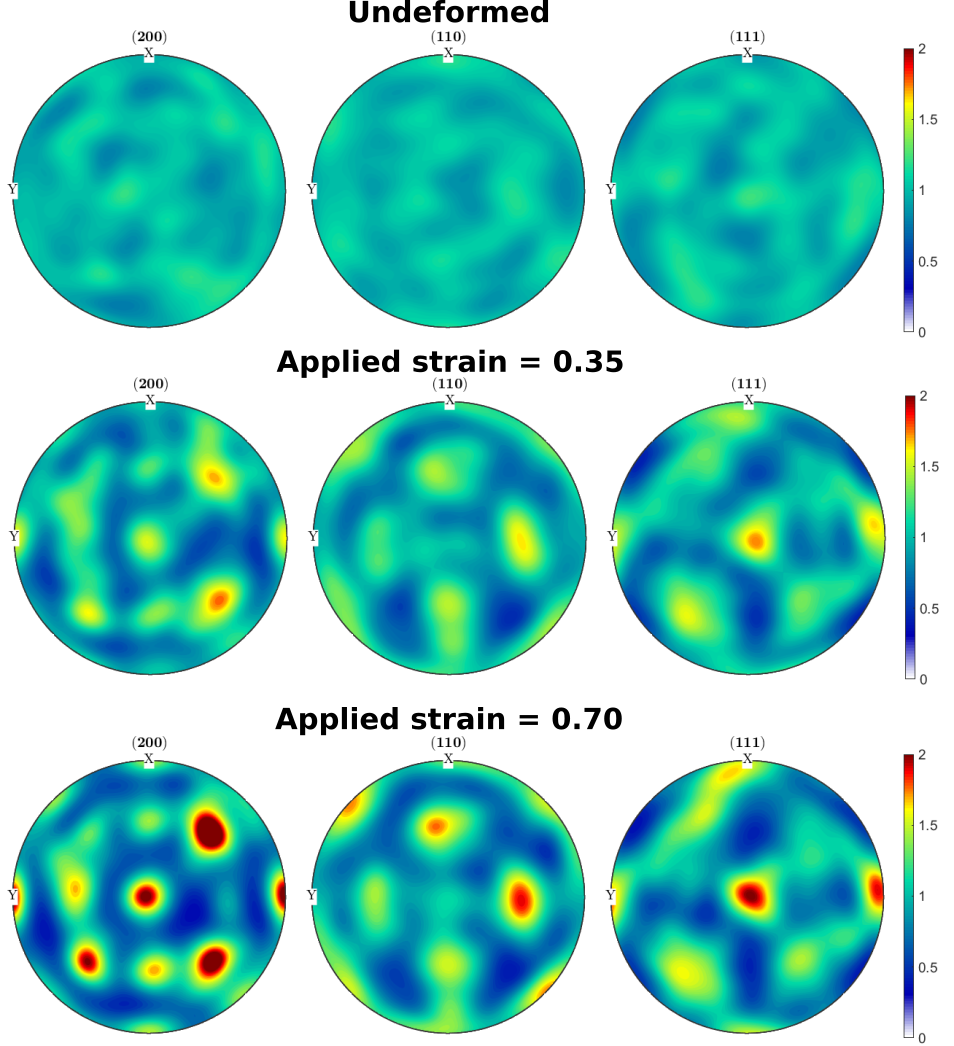}
	\caption{Initial and deformed (200), (110) and (111) pole figures of tantalum polycrystals, with 512 orientations, subjected to uniaxial compression at an applied strain rate of $5 \times 10^{3} /s$ at 298 K.}
	\label{fig:ta_texture}
\end{figure}

We have used the constitutive model to predict the texture evolution of tantalum during compressive loading. For these simulations, an initial random texture comprised of 512 orientations was used. The undeformed texture is shown in Figure \ref{fig:ta_texture} in terms of the (200), (110) and (111) pole figures. A cube-shaped simulation domain was meshed with 8 3D hexahedral finite elements per grain (total 4096 elements). Symmetric boundary conditions were applied as in the previous Sections. The simulation domain was subjected to uniaxial compression along the y-direction at a nominal strain rate of $5 \times 10^{3}$ /s up to 0.7 applied strain at 298 K. The texture evolution is plotted in terms of the (200), (110) and (111) pole figures after an applied strain of 0.35 and 0.7 in Figure \ref{fig:ta_texture}. It can be seen from these pole figures that while the undeformed texture is relatively random, with no observable pole intensities, some texture components start developing with applied strain. This texture evolution is qualitatively similar to the experimental texture given in \cite{kothari1998elasto, bingert1997texture}, with the development of stronger <001> and <111> texture components, as compared to the <110> component. 

\begin{figure}[!htbp]
    \centering
	\includegraphics[scale=0.55]{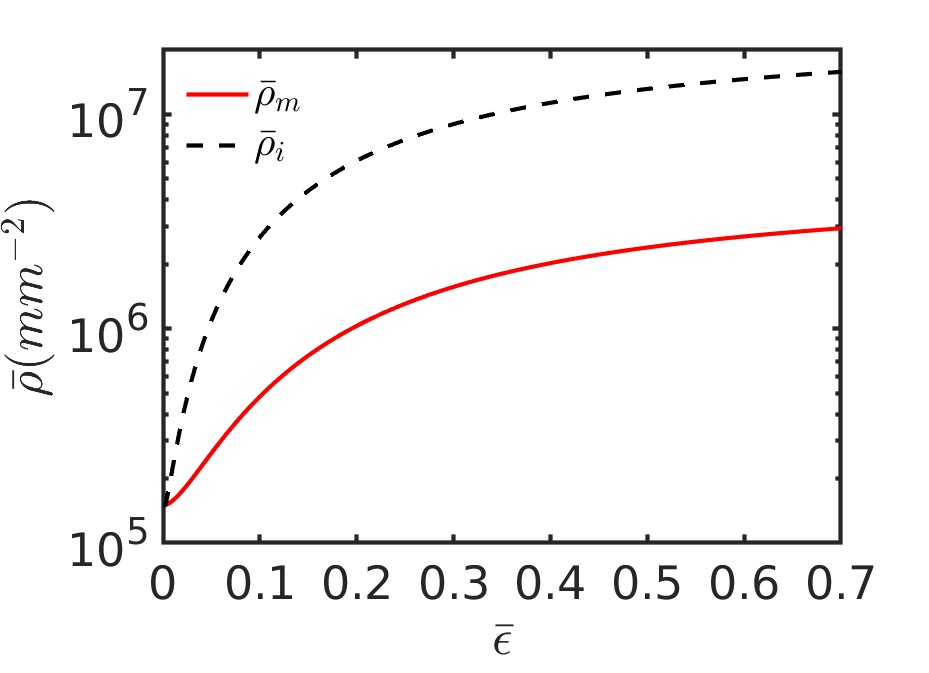}
	\caption{Evolution of slip system-averaged mobile and immobile dislocation densities as a function of applied strain for tantalum polycrystals subjected to uniaxial compression at an applied strain rate of $5 \times 10^{3} /s$ at 298 K.}
	\label{fig:ta_rho}
\end{figure}

The slip system-averaged mobile and immobile dislocation densities as a function of effective strain from the same simulation are plotted in Figure \ref{fig:ta_rho}. Note that the dislocation densities are plotted on the log scale. The initial dislocation density used in these simulations was representative of an annealed material. While there is a rapid increase in the dislocation densities during the initial stages, up to about 0.3 effective strain, the dislocation densities saturate subsequently. This correlates with the hardening response seen for tantalum in Figure \ref{fig:ta_stress_strain}. It can also be seen that the average immobile dislocation density is $\approx$ 5-10 times higher than the average mobile dislocation density at any stage. This prediction is in qualitative concurrence with prior studies \cite{bratov2015comparison}, where the immobile dislocation density dominates during the later stages of deformation. In our constitutive model, the dislocation multiplication parameter, $k^{\alpha}_{M}$, the dislocation immobilization parameter, $k^{\alpha}_{I}$ and the dynamic recovery parameter, $k^{\alpha}_{D}$, may be altered to obtain the desired ratio of mobile and immobile dislocation densities. 

\subsubsection{3D Simulations of Realistic Microstructures}

\begin{figure}[!htbp]
    \centering
	\includegraphics[scale=0.2]{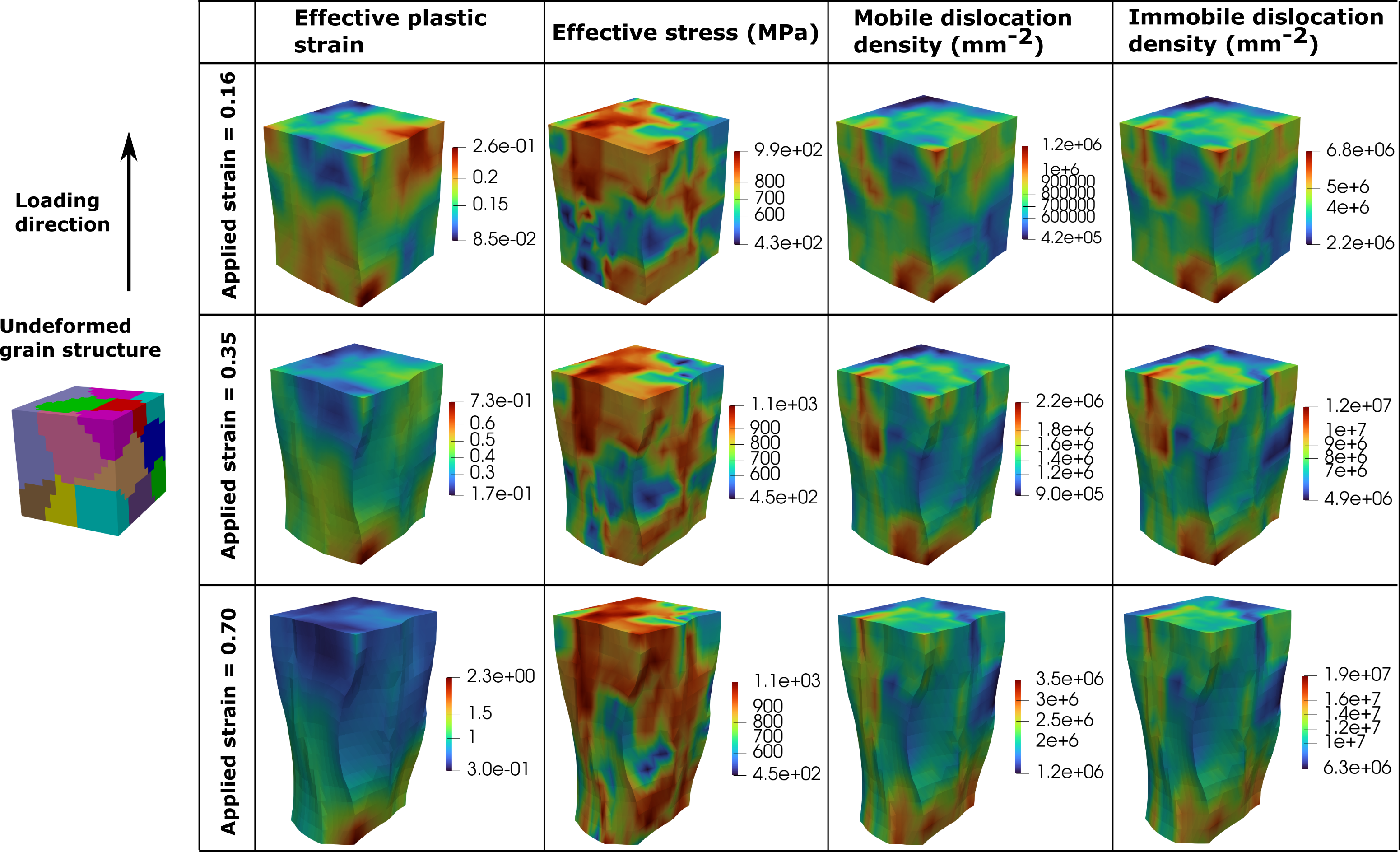}
	\caption{Contours of effective plastic strain, $\bar \epsilon^{p}$, effective stress, $\bar \sigma$, slip system-averaged mobile dislocation density, $\bar \rho_{m}$, and slip system-averaged immobile dislocation density, $\bar \rho_{i}$, at different applied strains during tensile deformation of a polycrystalline tantalum specimen. The initial grain structure was instantiated using a Voronoi tessellation algorithm in a cubic domain of side 15 $\mu m$. The tensile specimen was loaded using a strain rate of $10^{-3}$ /s at 114 K.}
	\label{fig:ta_3d_contours}
\end{figure}

While all the simulations in the previous Sections were performed using idealized cube-shaped grains, we demonstrate results with realistic 3D microstructures in this Section. For this, a synthetic microstructure was instantiated using an in-house Voronoi tessellation code. The cubic domain of 15 $\mu m$ side was meshed with 3D hexahedral elements having an element size of 1 $\mu m$, with 22 grains. The simulation domain had a total of 3375 elements, 4096 nodes and 12288 degrees of freedom. The grain structure of the undeformed microstructure is shown in Figure \ref{fig:ta_3d_contours}. As earlier, symmetric boundary conditions were used and displacement-controlled tensile loading was applied at a nominal strain rate of $1 \times 10^{-3}$ /s at 114 K.

\begin{figure}[!htbp]
    \centering
	\includegraphics[scale=0.55]{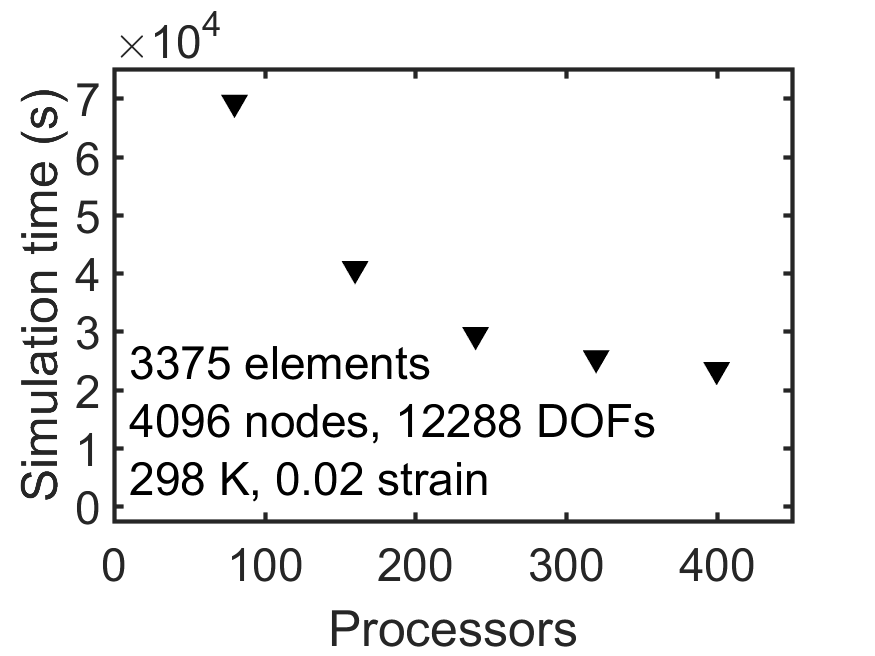}
	\caption{Simulation wall time as a function of the number of parallel processors for tensile deformation of a polycrystalline tantalum specimen, with 3,375 elements, 4,096 nodes and 12,288 degrees of freedom, loaded up to 0.02 nominal strain using a strain rate of $10^{-3}$ /s at 114 K.}
	\label{fig:ta_scaling}
\end{figure}

We first present the parallel scaling results for the simulations loaded till 0.02 nominal strain. For this, the same simulation was run on parallel processors ranging from 80 processors to 400 processors on the Param Sanganak supercomputer at IIT Kanpur, with Intel Xeon Platinum 8268 processors having 2.9 GHz clock speed and 4 GB memory per processor. The simulations were run using the implicit Newton solver in MOOSE, which we have generally found to provide the best convergence. The simulation time required for these simulations to reached 0.02 nominal strain is presented in Figure \ref{fig:ta_scaling}. A good parallel scaling is obtained up to 240 processors, after which the performance starts to saturate. Increasing the problem size might demonstrate better scaling with even higher number of processors. These parallel scaling capabilities are due to the inherent MOOSE architecture, and has been demonstrated to scale very well up to thousands of processors using a Jacobian-free Newton-Krylov (JFNK) solver \cite{permann2020moose}. However, the Newton solver has generally been found to provide better convergence in our simulations. In cases, where the memory load per processor is too heavy using the Newton solver, an explicit solver may be used for $\rho$-CP simulations.

Figure \ref{fig:ta_3d_contours} presents contours of the effective plastic strain, $\bar \epsilon^{p}$, effective stress, $\bar \sigma$, slip system-averaged mobile dislocation density, $\bar \rho_{m}$, and immobile dislocation density, $\bar \rho_{i}$, at different stages of tensile deformation for the same simulation loaded in tension up to 0.70 applied strain. It can be seen from these contours that heterogeneous deformation takes place between the grains to accommodate the imposed deformation. For example, very high strain localization ($\bar \epsilon^{p} = 2.3$) is observed near the bottom face at an applied strain of 0.7, while other regions have $\bar \epsilon^{p}$ as low as 0.3. Similar heterogeneity is also observed in the stress contours, especially in the regions near the grain interfaces. The mobile and immobile dislocation densities being a function of plastic strain, are generally high in the regions where the plastic strain is localized. It is also observed that the dislocation densities are high in the vicinity of grain interfaces, for example near the top left corner of the simulation domain. The shape of the deformed domain is also indicative of geometric localization due to necking. Overall, these results demonstrate the model's capability of predicting heterogeneous deformation in realistic microstructures.

\subsubsection{Simulation of EBSD Microstructure}

\begin{figure}[!htbp]
    \centering
	\includegraphics[scale=0.4]{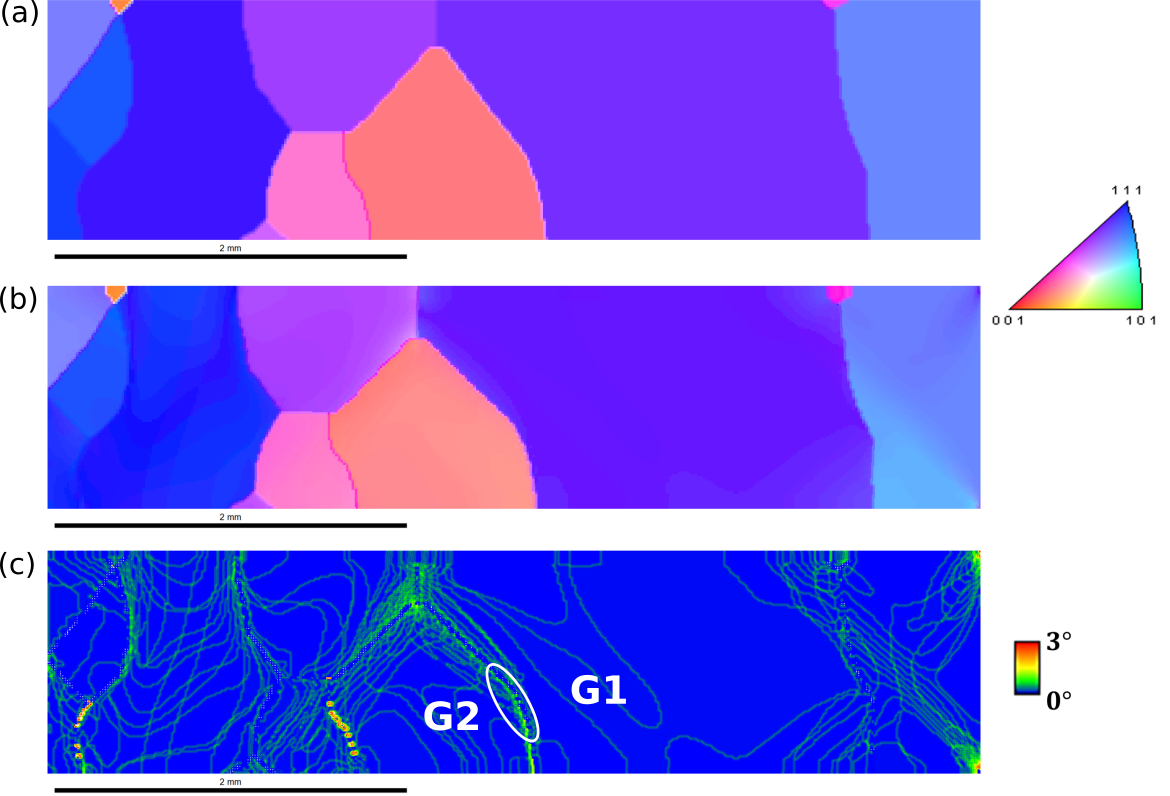}
	\caption{Inverse Pole Figure (IPF) map of (a) undeformed microstructure, (b) deformed microstructure, and (c) Kernel Average Misorientation (KAM) contour of the deformed microstructure of the tantalum oligocrystals after 0.10 applied strain using a strain rate of $10^{-3}$ /s at 298 K. Note that these quantities have been plotted on the undeformed mesh. Euler angles for the undeformed microstructure were obtained from \cite{lim2015quantitative}.}
	\label{fig:ta_KAM_contours}
\end{figure}

In this last example, we demonstrate the simulation of a tantalum oligocrystal microstructure obtained from Electron Back Scatter Diffraction (EBSD). For this simulation, we have digitized the EBSD microstructure of tantalum oligocrystals given in \cite{lim2015quantitative}. The microstructure, with dimensions of $5280 \times 1350 \mu m$, was mesh using 3D hexahedral elements having an element size of 15 $\mu m$. In this example simulation, only one layer of elements was considered into the plane, although it has been shown that considering more layers may increase the prediction accuracy \cite{lim2014grain, lim2015quantitative}. The Euler angles for the individual grains were taken directly from \cite{lim2015quantitative} and assigned to the digitized microstructure. The Inverse Pole Figure (IPF) map of the undeformed microstructure is shown in Figure \ref{fig:ta_KAM_contours} (a). The simulation domain had 32,123 elements, 65,136 nodes and 195,408 degrees of freedom. The back face of the simulation domain was constrained to move along direction z (into the plane), while the right face was constrained to move in directions x and y. Displacement-controlled tensile loading was applied on the left face along direction y at a nominal strain rate of $1 \times 10^{-3}$ /s up to 0.10 applied strain at 298 K. We note that the constitutive model for tantalum was calibrated to a different experimental response \cite{hoge1977temperature, nemat1997direct, kothari1998elasto}, while we are using the same here for predicting the deformation of this EBSD microstructure from \cite{lim2015quantitative}. This might indeed lead to some discrepancy in prediction, although the qualitative trends are expected to be similar.

\begin{figure}[!htbp]
    \centering
	\includegraphics[scale=0.3]{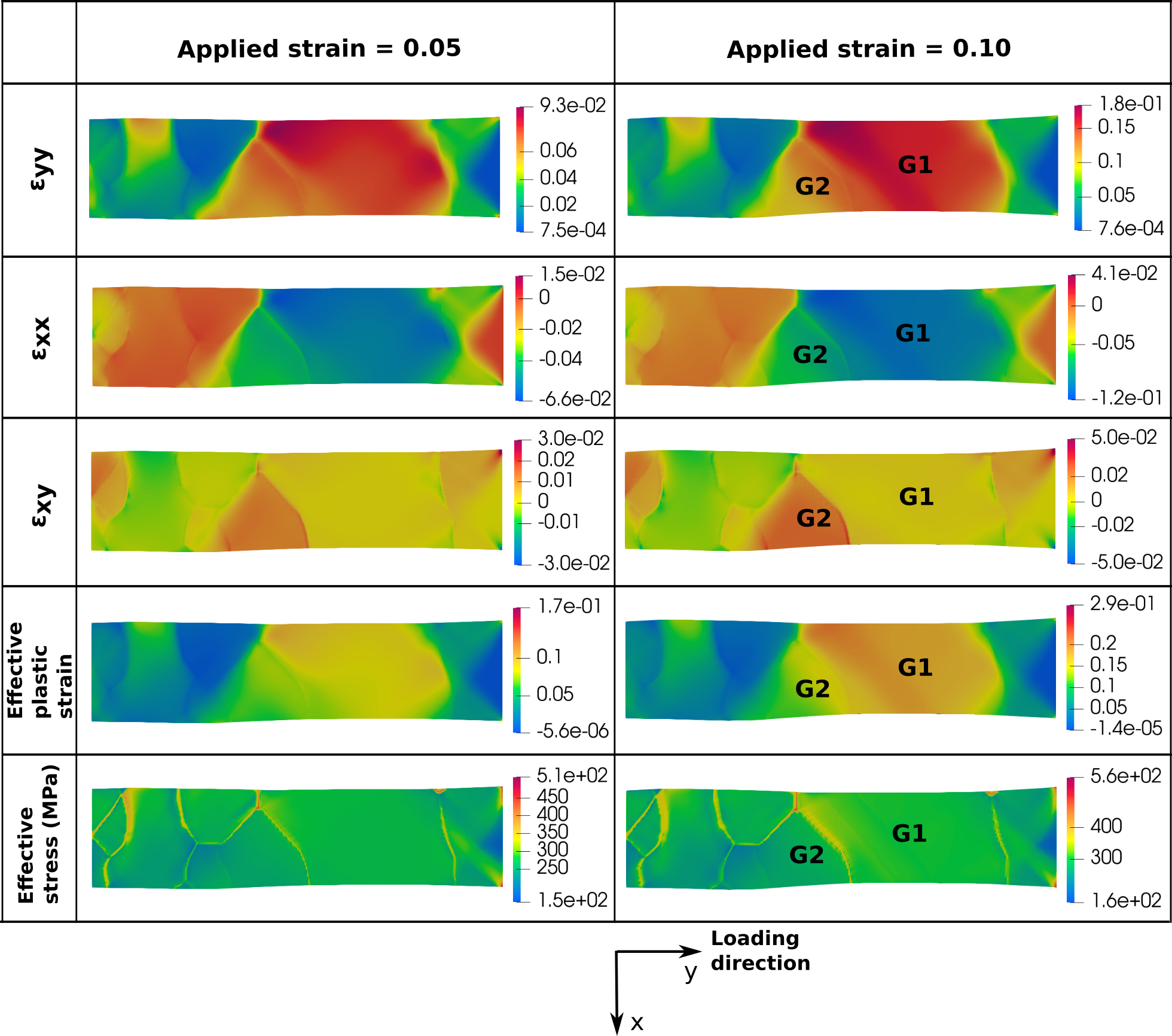}
	\caption{Contours of $\epsilon_{yy}$, $\epsilon_{xx}$, $\epsilon_{xy}$, effective plastic strain, $\bar \epsilon^{p}$, and effective stress, $\bar \sigma$, at 0.05 and 0.10 applied strains during tensile deformation of an EBSD microstructure of tantalum oligocrystals. The tensile specimen was loaded uniaxially along the direction y using a strain rate of $10^{-3}$ /s at 298 K. The initial EBSD microstructure was obtained by digitizing the microstructure and using the Euler angles given in \cite{lim2015quantitative}.}
	\label{fig:ta_EBSD_contours}
\end{figure}

Figure \ref{fig:ta_EBSD_contours} shows deformed contours of $\epsilon_{yy}$, $\epsilon_{xx}$, $\epsilon_{xy}$, effective plastic strain, $\bar \epsilon^{p}$, and effective stress, $\bar \sigma$, at 0.05 and 0.10 applied strain. The predicted deformed shape of the gauge region is qualitatively similar to the experimental observations in \cite{lim2015quantitative}. For example, there is a "neck" formation near the center of the gauge region. Moreover, similar to the experimental Digital Image Correlation (DIC) strain measurements \cite{lim2015quantitative}, the grain marked G1 was found to have the highest strain among all grains in the gauge. The experiments also showed strain localization at the boundary between grains marked G1 and G2 \cite{lim2015quantitative}, as is predicted here in the same regions (see $\epsilon_{xx}$ and $\epsilon_{xy}$ contours). The $\bar \epsilon^{p}$ contours show similar trends, while a high stress localization is generally observed at most grain interfaces in the $\bar \sigma$ contours. This is also discussed in terms of the misorientation development next.

The Euler angles of the deformed microstructure were used to plot the IPF map and the Kernel Average Misorientation (KAM) contours after 0.10 applied strain in Figures \ref{fig:ta_KAM_contours} (b) and (c), respectively. KAM, which is a local point-to-point misorientation measure, is generally attributed to the accommodation of heterogeneous deformation in regions with incompatible interfaces \cite{pai2022study}. As can be seen from the KAM contours, high misorientation is predicted at the same grain interface, where strain localizations were observed in our model predictions (Figure \ref{fig:ta_EBSD_contours}) and also in the experiments \cite{lim2015quantitative}. This is highlighted using the white elliptical marker between grains G1 and G2 in Figure \ref{fig:ta_KAM_contours}. Further note that all the boundaries / interfaces of the grain G2 show relatively higher misorientation development as compared to the other regions of the microstructure. This could be due to its location near the center of the gauge and also due to its intergranular heterogeneity with the neighboring grains. Failure in this tensile specimen may be expected to initiate at one of these interfaces. On the other hand, the IPF map does not show any significant evidence of grain rotation. This is expected given the relative grain sizes and the small strains applied.

Overall, we have demonstrated the model's ability to qualitatively predict the strain localization in an EBSD microstructure of tantalum oligocrystals. Although, we did not have the exact EBSD microstructure and considered only one layer of elements along the depth (neglecting any sub-surface grain effects), our model predicts the deformation contours with reasonable accuracy, including the regions of strain localization.

\section{Summary}

We have presented an open source, dislocation density based finite deformation crystal plasticity modeling framework, $\rho$-CP. The crystal plasticity model uses a Kocks-type thermally activated flow rule for modeling the temperature- and strain rate-effects on the crystallographic shearing rate. Mobile and immobile dislocation densities, as well as slip system-level backstress are used as internal state variables to represent the substructure evolution during plastic deformation. Further, twinning in hcp crystals has also been modeled. The framework relies on the fully implicit numerical integration of the crystal plasticity model and provides the updated stress and tangent stiffness tensor that can be passed to the finite element solver. $\rho$-CP has been integrated with the open source finite element framework, MOOSE, for performing crystal plasticity finite element (CPFE) simulations of deformation in metallic systems. Example applications have been demonstrated for predicting the anisotropic mechanical response of hcp magnesium single and polycrystals, strain rate effects and slip system-level backstress induced hardening during cyclic loading in polycrystalline fcc OFHC copper, and temperature- and strain rate-effects on the deformation of polycrystalline bcc tantalum. The model has also been used to run CPFE simulations on realistic microstructures including EBSD microstructures of tantalum oligocrystals to predict the misorientation development during tensile deformation. Overall, we have demonstrated the model's ability to predict both macroscopic mechanical properties as well as local microstructure evolution.

The framework presented here is general enough to allow consideration for material- and microstructure-specific strengthening mechanisms for various metallic systems. Further, on-the-run deformation mode and material property assignment allow the user to run simulations for different metals and alloys, with minimum code development or implementation. Integration of $\rho$-CP with MOOSE allows leveraging features, such as parallelization on hundreds of processors, multi-physics coupling, as well as restarting terminated simulations, which are inherently part of the MOOSE architecture.

$\rho$-CP can serve as a tool for both new users as well as experienced crystal plasticity developers for simulating deformation in polycrystalline ensembles. 

The C++ source codes and input files for the example simulations are shared in the github repository: https://github.com/apatra6/rhocp

\section*{Acknowledgments}
AP acknowledges partial funding received for this work from the Department of Science and Technology (DST) - Science and Engineering Research Board (SERB), India under grant number: CRG/2020/000593. The support and the resources provided by PARAM Sanganak under the National Supercomputing Mission, Government of India, at the Indian Institute of Technology, Kanpur are gratefully acknowledged for running the parallel scaling simulations presented in Figure \ref{fig:ta_scaling}.

\section*{Appendix: List of examples in the $\rho$-CP repository}
\begin{table}[!htbp]
   \footnotesize
   \begin{center}
     \caption{Details of example simulations in the $\rho$-CP repository.}
     \begin{tabular}{c c c}
      \hline
      Name & Results & Path \\
      \hline
      Mg single crystal plane strain compression & Figures \ref{fig:mg_sx_stress_strain} and \ref{fig:mg_sx_slip_activity} & \texttt{examples/magnesium/SX} \\
      Mg polycrystal plane strain compression & Figures \ref{fig:mg_px_stress_strain} and \ref{fig:mg_px_slip_activity} & \texttt{examples/magnesium/PX} \\
      OFHC Cu strain rate effect & Figure \ref{fig:cu_strain_rate} & \texttt{examples/copper/strain\_rate\_effect} \\
      OFHC Cu cyclic deformation & Figure \ref{fig:cu_cyclic} & \texttt{examples/copper/cyclic\_test} \\
      Ta yield stress temperature effect & Figure \ref{fig:ta_ys}(a) & \texttt{examples/tantalum/temperature\_effect} \\
      Ta yield stress strain rate effect & Figure \ref{fig:ta_ys}(b) & \texttt{examples/tantalum/strain\_rate\_effect} \\
      Ta uniaxial compression (64 grains) & Figure \ref{fig:ta_stress_strain} & \texttt{examples/tantalum/temperature\_effect/compression\_64} \\
      Ta uniaxial compression (512 grains) & Figures \ref{fig:ta_texture} and \ref{fig:ta_rho} & \texttt{examples/tantalum/temperature\_effect/compression\_512} \\
      Ta uniaxial tension & Figure \ref{fig:ta_3d_contours} & \texttt{examples/tantalum/3d\_pxtal} \\
      Ta EBSD simulation & Figures \ref{fig:ta_EBSD_contours} and \ref{fig:ta_KAM_contours} & \texttt{examples/tantalum/EBSD\_simulation} \\
      \hline
      \end{tabular}
   \end{center}
   \label{examples_table}
\end{table}

\bibliographystyle{unsrt}  
\bibliography{references}

\begin{thebibliography}{100}

\bibitem{mcdowell2008viscoplasticity}
David~L McDowell.
\newblock Viscoplasticity of heterogeneous metallic materials.
\newblock {\em Materials Science and Engineering: R: Reports}, 62(3):67--123,
  2008.

\bibitem{mcdowell2010perspective}
David~L McDowell.
\newblock A perspective on trends in multiscale plasticity.
\newblock {\em International Journal of Plasticity}, 26(9):1280--1309, 2010.

\bibitem{roters2010overview}
Franz Roters, Philip Eisenlohr, Luc Hantcherli, Denny~Dharmawan Tjahjanto,
  Thomas~R Bieler, and Dierk Raabe.
\newblock Overview of constitutive laws, kinematics, homogenization and
  multiscale methods in crystal plasticity finite-element modeling: Theory,
  experiments, applications.
\newblock {\em Acta Materialia}, 58(4):1152--1211, 2010.

\bibitem{repetto1997micromechanical}
EA~Repetto and M~Ortiz.
\newblock A micromechanical model of cyclic deformation and fatigue-crack
  nucleation in fcc single crystals.
\newblock {\em Acta Materialia}, 45(6):2577--2595, 1997.

\bibitem{kalidindi1998incorporation}
Surya~R Kalidindi.
\newblock Incorporation of deformation twinning in crystal plasticity models.
\newblock {\em Journal of the Mechanics and Physics of Solids}, 46(2):267--290,
  1998.

\bibitem{kothari1998elasto}
M~Kothari and L~Anand.
\newblock Elasto-viscoplastic constitutive equations for polycrystalline
  metals: application to tantalum.
\newblock {\em Journal of the Mechanics and Physics of Solids}, 46(1):51--83,
  1998.

\bibitem{tjahjanto2008crystallographically}
DD~Tjahjanto, S~Turteltaub, and ASJ Suiker.
\newblock Crystallographically based model for transformation-induced
  plasticity in multiphase carbon steels.
\newblock {\em Continuum Mechanics and Thermodynamics}, 19(7):399--422, 2008.

\bibitem{austin2011dislocation}
Ryan~A Austin and David~L McDowell.
\newblock A dislocation-based constitutive model for viscoplastic deformation
  of fcc metals at very high strain rates.
\newblock {\em International Journal of Plasticity}, 27(1):1--24, 2011.

\bibitem{beyerlein2003modeling}
IJ~Beyerlein, RA~Lebensohn, and CN~Tome.
\newblock Modeling texture and microstructural evolution in the equal channel
  angular extrusion process.
\newblock {\em Materials Science and Engineering: A}, 345(1-2):122--138, 2003.

\bibitem{li2005crystal}
Saiyi Li, Surya~R Kalidindi, and Irene~J Beyerlein.
\newblock A crystal plasticity finite element analysis of texture evolution in
  equal channel angular extrusion.
\newblock {\em Materials Science and Engineering: A}, 410:207--212, 2005.

\bibitem{jia2012non}
N~Jia, F~Roters, P~Eisenlohr, C~Kords, and D~Raabe.
\newblock Non-crystallographic shear banding in crystal plasticity fem
  simulations: Example of texture evolution in $\alpha$-brass.
\newblock {\em Acta Materialia}, 60(3):1099--1115, 2012.

\bibitem{zhang2016virtual}
Haiming Zhang, Martin Diehl, Franz Roters, and Dierk Raabe.
\newblock A virtual laboratory using high resolution crystal plasticity
  simulations to determine the initial yield surface for sheet metal forming
  operations.
\newblock {\em International Journal of Plasticity}, 80:111--138, 2016.

\bibitem{kysar2007high}
Jeffrey~W Kysar, Yong~X Gan, Timothy~L Morse, Xi~Chen, and Milton~E Jones.
\newblock High strain gradient plasticity associated with wedge indentation
  into face-centered cubic single crystals: geometrically necessary dislocation
  densities.
\newblock {\em Journal of the Mechanics and Physics of Solids},
  55(7):1554--1573, 2007.

\bibitem{zhang2010simulation}
M~Zhang, F~Bridier, P~Villechaise, J~Mendez, and DL~McDowell.
\newblock Simulation of slip band evolution in duplex ti--6al--4v.
\newblock {\em Acta Materialia}, 58(3):1087--1096, 2010.

\bibitem{guery2016slip}
Adrien Guery, Fran{\c{c}}ois Hild, F{\'e}lix Latourte, and St{\'e}phane Roux.
\newblock Slip activities in polycrystals determined by coupling dic
  measurements with crystal plasticity calculations.
\newblock {\em International Journal of Plasticity}, 81:249--266, 2016.

\bibitem{guan2017crystal}
Yongjun Guan, Bo~Chen, Jinwen Zou, T~Ben Britton, Jun Jiang, and Fionn~PE
  Dunne.
\newblock Crystal plasticity modelling and hr-dic measurement of slip
  activation and strain localization in single and oligo-crystal ni alloys
  under fatigue.
\newblock {\em International Journal of Plasticity}, 88:70--88, 2017.

\bibitem{ganesan2021effects}
Sriram Ganesan, Mohammadreza Yaghoobi, Alan Githens, Zhe Chen, Samantha Daly,
  John~E Allison, and Veera Sundararaghavan.
\newblock The effects of heat treatment on the response of we43 mg alloy:
  crystal plasticity finite element simulation and sem-dic experiment.
\newblock {\em International Journal of Plasticity}, 137:102917, 2021.

\bibitem{zhang2012phenomenological}
Jing Zhang and Shailendra~P Joshi.
\newblock Phenomenological crystal plasticity modeling and detailed
  micromechanical investigations of pure magnesium.
\newblock {\em Journal of the Mechanics and Physics of Solids}, 60(5):945--972,
  2012.

\bibitem{lim2014grain}
Hojun Lim, JD~Carroll, Corbett~Chandler Battaile, TE~Buchheit, BL~Boyce, and
  CR~Weinberger.
\newblock Grain-scale experimental validation of crystal plasticity finite
  element simulations of tantalum oligocrystals.
\newblock {\em International Journal of Plasticity}, 60:1--18, 2014.

\bibitem{bittencourt2019interpretation}
Eduardo Bittencourt.
\newblock Interpretation of the size effects in micropillar compression by a
  strain gradient crystal plasticity theory.
\newblock {\em International Journal of Plasticity}, 116:280--296, 2019.

\bibitem{mcdowell2010microstructure}
DL~McDowell and FPE Dunne.
\newblock Microstructure-sensitive computational modeling of fatigue crack
  formation.
\newblock {\em International Journal of Fatigue}, 32(9):1521--1542, 2010.

\bibitem{stopka2022simulated}
Krzysztof~S Stopka, Mohammadreza Yaghoobi, John~E Allison, and David~L
  McDowell.
\newblock Simulated effects of sample size and grain neighborhood on the
  modeling of extreme value fatigue response.
\newblock {\em Acta Materialia}, 224:117524, 2022.

\bibitem{lebensohn1993self}
Ricardo~A Lebensohn and CN~Tom{\'e}.
\newblock A self-consistent anisotropic approach for the simulation of plastic
  deformation and texture development of polycrystals: application to zirconium
  alloys.
\newblock {\em Acta metallurgica et materialia}, 41(9):2611--2624, 1993.

\bibitem{van2006multiscale}
Paul Van~Houtte, Anand~Krishna Kanjarla, Albert Van~Bael, Marc Seefeldt, and
  Laurent Delannay.
\newblock Multiscale modelling of the plastic anisotropy and deformation
  texture of polycrystalline materials.
\newblock {\em European Journal of Mechanics-A/Solids}, 25(4):634--648, 2006.

\bibitem{lebensohn2012elasto}
Ricardo~A Lebensohn, Anand~K Kanjarla, and Philip Eisenlohr.
\newblock An elasto-viscoplastic formulation based on fast fourier transforms
  for the prediction of micromechanical fields in polycrystalline materials.
\newblock {\em International Journal of Plasticity}, 32:59--69, 2012.

\bibitem{roters2019damask}
Franz Roters, Martin Diehl, Pratheek Shanthraj, Philip Eisenlohr, C~Reuber,
  Su~Leen Wong, Tias Maiti, Alireza Ebrahimi, Thomas Hochrainer, H-O Fabritius,
  et~al.
\newblock Damask--the d{\"u}sseldorf advanced material simulation kit for
  modeling multi-physics crystal plasticity, thermal, and damage phenomena from
  the single crystal up to the component scale.
\newblock {\em Computational Materials Science}, 158:420--478, 2019.

\bibitem{lee1969elastic}
Erastus~H Lee.
\newblock Elastic-plastic deformation at finite strains.
\newblock {\em Journal of Applied Mechanics}, 36:1--6, 1969.

\bibitem{asaro1977strain}
Robert~J Asaro and JR0375 Rice.
\newblock Strain localization in ductile single crystals.
\newblock {\em Journal of the Mechanics and Physics of Solids}, 25(5):309--338,
  1977.

\bibitem{peirce1982analysis}
D~Peirce, RJ~Asaro, and A~Needleman.
\newblock An analysis of nonuniform and localized deformation in ductile single
  crystals.
\newblock {\em Acta metallurgica}, 30(6):1087--1119, 1982.

\bibitem{asaro1985overview}
Robert~J Asaro and Alan Needleman.
\newblock Overview no. 42 texture development and strain hardening in rate
  dependent polycrystals.
\newblock {\em Acta metallurgica}, 33(6):923--953, 1985.

\bibitem{hutchinson1976bounds}
John~Woodside Hutchinson.
\newblock Bounds and self-consistent estimates for creep of polycrystalline
  materials.
\newblock {\em Proceedings of the Royal Society of London. A. Mathematical and
  Physical Sciences}, 348(1652):101--127, 1976.

\bibitem{taylor1934mechanism}
Geoffrey~Ingram Taylor.
\newblock The mechanism of plastic deformation of crystals. part
  i.—theoretical.
\newblock {\em Proceedings of the Royal Society of London. Series A, Containing
  Papers of a Mathematical and Physical Character}, 145(855):362--387, 1934.

\bibitem{estrin1996dislocation}
Yuri Estrin.
\newblock Dislocation-density-related constitutive modeling.
\newblock {\em Unified constitutive laws of plastic deformation}, 1:69--106,
  1996.

\bibitem{zikry1996inelastic}
MA~Zikry and M~Kao.
\newblock Inelastic microstructural failure mechanisms in crystalline materials
  with high angle grain boundaries.
\newblock {\em Journal of the Mechanics and Physics of Solids},
  44(11):1765--1798, 1996.

\bibitem{kocks2003physics}
UF~Kocks and H~Mecking.
\newblock Physics and phenomenology of strain hardening: the fcc case.
\newblock {\em Progress in materials science}, 48(3):171--273, 2003.

\bibitem{ma2004constitutive}
A~Ma and F~Roters.
\newblock A constitutive model for fcc single crystals based on dislocation
  densities and its application to uniaxial compression of aluminium single
  crystals.
\newblock {\em Acta Materialia}, 52(12):3603--3612, 2004.

\bibitem{wong2016crystal}
Su~Leen Wong, Manjunatha Madivala, Ulrich Prahl, Franz Roters, and Dierk Raabe.
\newblock A crystal plasticity model for twinning-and transformation-induced
  plasticity.
\newblock {\em Acta Materialia}, 118:140--151, 2016.

\bibitem{feng2022crystal}
Zhangxi Feng, Reeju Pokharel, Sven~C Vogel, Ricardo~A Lebensohn, Darren Pagan,
  Eloisa Zepeda-Alarcon, Bj{\o}rn Clausen, Ramon Martinez, George~T Gray~III,
  and Marko Knezevic.
\newblock Crystal plasticity modeling of strain-induced martensitic
  transformations to predict strain rate and temperature sensitive behavior of
  304 l steels: Applications to tension, compression, torsion, and impact.
\newblock {\em International Journal of Plasticity}, 156:103367, 2022.

\bibitem{arsenlis1999crystallographic}
A~Arsenlis and DM~Parks.
\newblock Crystallographic aspects of geometrically-necessary and
  statistically-stored dislocation density.
\newblock {\em Acta Materialia}, 47(5):1597--1611, 1999.

\bibitem{gurtin2002gradient}
Morton~E Gurtin.
\newblock A gradient theory of single-crystal viscoplasticity that accounts for
  geometrically necessary dislocations.
\newblock {\em Journal of the Mechanics and Physics of Solids}, 50(1):5--32,
  2002.

\bibitem{evers2004non}
LP~Evers, WAM Brekelmans, and MGD1115 Geers.
\newblock Non-local crystal plasticity model with intrinsic ssd and gnd
  effects.
\newblock {\em Journal of the Mechanics and Physics of Solids},
  52(10):2379--2401, 2004.

\bibitem{mayeur2011dislocation}
Jason~R Mayeur, David~L McDowell, and Douglas~J Bammann.
\newblock Dislocation-based micropolar single crystal plasticity: Comparison of
  multi-and single criterion theories.
\newblock {\em Journal of the Mechanics and Physics of Solids}, 59(2):398--422,
  2011.

\bibitem{dunne2012crystal}
FPE Dunne, R~Kiwanuka, and AJ~Wilkinson.
\newblock Crystal plasticity analysis of micro-deformation, lattice rotation
  and geometrically necessary dislocation density.
\newblock {\em Proceedings of the Royal Society A: Mathematical, Physical and
  Engineering Sciences}, 468(2145):2509--2531, 2012.

\bibitem{pai2022study}
Namit Pai, Aditya Prakash, Indradev Samajdar, and Anirban Patra.
\newblock Study of grain boundary orientation gradients through combined
  experiments and strain gradient crystal plasticity modeling.
\newblock {\em International Journal of Plasticity}, 156:103360, 2022.

\bibitem{yaghoobi2021crystal}
Mohammadreza Yaghoobi, George~Z Voyiadjis, and Veera Sundararaghavan.
\newblock Crystal plasticity simulation of magnesium and its alloys: A review
  of recent advances.
\newblock {\em Crystals}, 11(4):435, 2021.

\bibitem{giannozzi2009quantum}
Paolo Giannozzi, Stefano Baroni, Nicola Bonini, Matteo Calandra, Roberto Car,
  Carlo Cavazzoni, Davide Ceresoli, Guido~L Chiarotti, Matteo Cococcioni,
  Ismaila Dabo, et~al.
\newblock Quantum espresso: a modular and open-source software project for
  quantum simulations of materials.
\newblock {\em Journal of physics: Condensed matter}, 21(39):395502, 2009.

\bibitem{thompson2022lammps}
Aidan~P Thompson, H~Metin Aktulga, Richard Berger, Dan~S Bolintineanu,
  W~Michael Brown, Paul~S Crozier, Pieter~J in't Veld, Axel Kohlmeyer, Stan~G
  Moore, Trung~Dac Nguyen, et~al.
\newblock Lammps-a flexible simulation tool for particle-based materials
  modeling at the atomic, meso, and continuum scales.
\newblock {\em Computer Physics Communications}, 271:108171, 2022.

\bibitem{arsenlis2007enabling}
Athanasios Arsenlis, Wei Cai, Meijie Tang, Moono Rhee, Tomas Oppelstrup, Gregg
  Hommes, Tom~G Pierce, and Vasily~V Bulatov.
\newblock Enabling strain hardening simulations with dislocation dynamics.
\newblock {\em Modelling and Simulation in Materials Science and Engineering},
  15(6):553, 2007.

\bibitem{xu2018pycac}
Shuozhi Xu, Thomas~G Payne, Hao Chen, Yongchao Liu, Liming Xiong, Youping Chen,
  and David~L McDowell.
\newblock Pycac: The concurrent atomistic-continuum simulation environment.
\newblock {\em Journal of Materials Research}, 33(7):857--871, 2018.

\bibitem{tonks2012object}
Michael~R Tonks, Derek Gaston, Paul~C Millett, David Andrs, and Paul Talbot.
\newblock An object-oriented finite element framework for multiphysics phase
  field simulations.
\newblock {\em Computational Materials Science}, 51(1):20--29, 2012.

\bibitem{dewitt2020prisms}
Stephen DeWitt, Shiva Rudraraju, David Montiel, W~Beck Andrews, and Katsuyo
  Thornton.
\newblock Prisms-pf: A general framework for phase-field modeling with a
  matrix-free finite element method.
\newblock {\em npj Computational Materials}, 6(1):1--12, 2020.

\bibitem{microsim}
Tanmay Datta, Dasari Mohan, Ajay Sagar, Saurav Shenoy, Swapnil Bhure, Abhishek
  Kalokhe, Nasir Attar, Swaroop Sampanand, MP~Gururajan, Venkatesh Shenoi,
  Vaishali Shah, Saswata Bhattacharyya, Gandham Phanikumar, and Abhik
  Choudhury.
\newblock "microsim: A high-performance phase-field software based on cpu and
  gpu implementations.
\newblock \url{https://microsim.co.in/}.
\newblock Accessed: 2023-02-05.

\bibitem{huang1991user}
Yonggang Huang.
\newblock {\em A user-material subroutine incroporating single crystal
  plasticity in the ABAQUS finite element program}.
\newblock Harvard Univ. Cambridge, MA, 1991.

\bibitem{yaghoobi2019prisms}
Mohammadreza Yaghoobi, Sriram Ganesan, Srihari Sundar, Aaditya Lakshmanan,
  Shiva Rudraraju, John~E Allison, and Veera Sundararaghavan.
\newblock Prisms-plasticity: An open-source crystal plasticity finite element
  software.
\newblock {\em Computational Materials Science}, 169:109078, 2019.

\bibitem{brough2017materials}
David~B Brough, Daniel Wheeler, and Surya~R Kalidindi.
\newblock Materials knowledge systems in python—a data science framework for
  accelerated development of hierarchical materials.
\newblock {\em Integrating materials and manufacturing innovation},
  6(1):36--53, 2017.

\bibitem{permann2020moose}
Cody~J Permann, Derek~R Gaston, David Andr{\v{s}}, Robert~W Carlsen, Fande
  Kong, Alexander~D Lindsay, Jason~M Miller, John~W Peterson, Andrew~E
  Slaughter, Roy~H Stogner, et~al.
\newblock Moose: Enabling massively parallel multiphysics simulation.
\newblock {\em SoftwareX}, 11:100430, 2020.

\bibitem{code_aster}
Code Aster.
\newblock Structures and thermomechanics analysis for studies and research.
\newblock \url{http://www.code-aster.org/}, 2022.

\bibitem{alnaes2015fenics}
Martin Aln{\ae}s, Jan Blechta, Johan Hake, August Johansson, Benjamin Kehlet,
  Anders Logg, Chris Richardson, Johannes Ring, Marie~E Rognes, and Garth~N
  Wells.
\newblock The fenics project version 1.5.
\newblock {\em Archive of Numerical Software}, 3(100), 2015.

\bibitem{jasak2007openfoam}
Hrvoje Jasak, Aleksandar Jemcov, Zeljko Tukovic, et~al.
\newblock Openfoam: A c++ library for complex physics simulations.
\newblock In {\em International workshop on coupled methods in numerical
  dynamics}, volume 1000, pages 1--20. IUC Dubrovnik Croatia, 2007.

\bibitem{MR3043640}
F.~Hecht.
\newblock New development in freefem++.
\newblock {\em J. Numer. Math.}, 20(3-4):251--265, 2012.

\bibitem{badia2018fempar}
Santiago Badia, Alberto~F Mart{\'\i}n, and Javier Principe.
\newblock Fempar: An object-oriented parallel finite element framework.
\newblock {\em Archives of Computational Methods in Engineering},
  25(2):195--271, 2018.

\bibitem{stukowski2009visualization}
Alexander Stukowski.
\newblock Visualization and analysis of atomistic simulation data with
  ovito--the open visualization tool.
\newblock {\em Modelling and simulation in materials science and engineering},
  18(1):015012, 2009.

\bibitem{quey2011large}
Romain Quey, PR~Dawson, and Fabrice Barbe.
\newblock Large-scale 3d random polycrystals for the finite element method:
  Generation, meshing and remeshing.
\newblock {\em Computer Methods in Applied Mechanics and Engineering},
  200(17-20):1729--1745, 2011.

\bibitem{groeber2014dream}
Michael~A Groeber and Michael~A Jackson.
\newblock Dream. 3d: a digital representation environment for the analysis of
  microstructure in 3d.
\newblock {\em Integrating materials and manufacturing innovation},
  3(1):56--72, 2014.

\bibitem{ahrens2005paraview}
James Ahrens, Berk Geveci, and Charles Law.
\newblock Paraview: An end-user tool for large data visualization.
\newblock {\em The visualization handbook}, 717(8), 2005.

\bibitem{thool2020role}
Khushahal Thool, Anirban Patra, David Fullwood, KV~Mani Krishna, Dinesh
  Srivastava, and Indradev Samajdar.
\newblock The role of crystallographic orientations on heterogeneous
  deformation in a zirconium alloy: a combined experimental and modeling study.
\newblock {\em International Journal of Plasticity}, 133:102785, 2020.

\bibitem{pokharel2019analysis}
Reeju Pokharel, Anirban Patra, Donald~W Brown, Bj{\o}rn Clausen, Sven~C Vogel,
  and George~T Gray~III.
\newblock An analysis of phase stresses in additively manufactured 304l
  stainless steel using neutron diffraction measurements and crystal plasticity
  finite element simulations.
\newblock {\em International Journal of Plasticity}, 121:201--217, 2019.

\bibitem{patra2012crystal}
Anirban Patra and David~L McDowell.
\newblock Crystal plasticity-based constitutive modelling of irradiated bcc
  structures.
\newblock {\em Philosophical Magazine}, 92(7):861--887, 2012.

\bibitem{patra2015void}
Anirban Patra and David~L McDowell.
\newblock A void nucleation and growth based damage framework to model failure
  initiation ahead of a sharp notch in irradiated bcc materials.
\newblock {\em Journal of the Mechanics and Physics of Solids}, 74:111--135,
  2015.

\bibitem{patra2016crystal}
Anirban Patra and David~L McDowell.
\newblock Crystal plasticity investigation of the microstructural factors
  influencing dislocation channeling in a model irradiated bcc material.
\newblock {\em Acta Materialia}, 110:364--376, 2016.

\bibitem{patra2014constitutive}
Anirban Patra, Ting Zhu, and David~L McDowell.
\newblock Constitutive equations for modeling non-schmid effects in single
  crystal bcc-fe at low and ambient temperatures.
\newblock {\em International Journal of Plasticity}, 59:1--14, 2014.

\bibitem{ranjan2021crystal}
Devraj Ranjan, Sankar Narayanan, Kai Kadau, and Anirban Patra.
\newblock Crystal plasticity modeling of non-schmid yield behavior: from ni3al
  single crystals to ni-based superalloys.
\newblock {\em Modelling and Simulation in Materials Science and Engineering},
  29(5):055005, 2021.

\bibitem{geers2014coupled}
MGD Geers, Maeva Cottura, Benoit Appolaire, Esteban~P Busso, Samuel Forest, and
  Aur{\'e}lien Villani.
\newblock Coupled glide-climb diffusion-enhanced crystal plasticity.
\newblock {\em Journal of the Mechanics and Physics of Solids}, 70:136--153,
  2014.

\bibitem{chaudhary2022crystal}
Suketa Chaudhary, PJ~Guruprasad, and Anirban Patra.
\newblock Crystal plasticity constitutive modeling of tensile, creep and cyclic
  deformation in single crystal ni-based superalloys.
\newblock {\em Mechanics of Materials}, 174:104474, 2022.

\bibitem{kocks1975thermodynamics}
UF~Kocks, A.S. Argon, and MF~Ashby.
\newblock Thermodynamics and kinetics of slip.
\newblock {\em Progress in Materials Science}, 19:1, 1975.

\bibitem{meyers2001onset}
MA~Meyers, O~V{\"o}hringer, and VA~Lubarda.
\newblock The onset of twinning in metals: a constitutive description.
\newblock {\em Acta Materialia}, 49(19):4025--4039, 2001.

\bibitem{beyerlein2008dislocation}
IJ~Beyerlein and CN~Tom{\'e}.
\newblock A dislocation-based constitutive law for pure zr including
  temperature effects.
\newblock {\em International Journal of Plasticity}, 24(5):867--895, 2008.

\bibitem{oppedal2012effect}
AL~Oppedal, H~El~Kadiri, CN~Tom{\'e}, GC~Kaschner, Sven~C Vogel, JC~Baird, and
  MF~Horstemeyer.
\newblock Effect of dislocation transmutation on modeling hardening mechanisms
  by twinning in magnesium.
\newblock {\em International Journal of Plasticity}, 30:41--61, 2012.

\bibitem{cheng2017crystal}
Jiahao Cheng and Somnath Ghosh.
\newblock Crystal plasticity finite element modeling of discrete twin evolution
  in polycrystalline magnesium.
\newblock {\em Journal of the Mechanics and Physics of Solids}, 99:512--538,
  2017.

\bibitem{abdolvand2020nucleation}
Hamidreza Abdolvand, Karim Louca, Charles Mareau, Marta Majkut, and Jonathan
  Wright.
\newblock On the nucleation of deformation twins at the early stages of
  plasticity.
\newblock {\em Acta Materialia}, 196:733--746, 2020.

\bibitem{hall1951deformation}
EO~Hall.
\newblock The deformation and ageing of mild steel: Iii discussion of results.
\newblock {\em Proceedings of the Physical Society. Section B}, 64(9):747,
  1951.

\bibitem{petch1953cleavage}
NJ~Petch.
\newblock The cleavage strength of polycrystals.
\newblock {\em Journal of the Iron and Steel institute}, 174:25--28, 1953.

\bibitem{fleischer1963substitutional}
Robert~L Fleischer.
\newblock Substitutional solution hardening.
\newblock {\em Acta metallurgica}, 11(3):203--209, 1963.

\bibitem{labusch1970statistical}
Rea Labusch.
\newblock A statistical theory of solid solution hardening.
\newblock {\em physica status solidi (b)}, 41(2):659--669, 1970.

\bibitem{nabarro1997theoretical}
FRN Nabarro.
\newblock Theoretical and experimental estimates of the peierls stress.
\newblock {\em Philosophical Magazine A}, 75(3):703--711, 1997.

\bibitem{christian1995deformation}
John~Wyrill Christian and Subhash Mahajan.
\newblock Deformation twinning.
\newblock {\em Progress in materials science}, 39(1-2):1--157, 1995.

\bibitem{graff2007yielding}
St{\'e}phane Graff, Wolfgang Brocks, and Dirk Steglich.
\newblock Yielding of magnesium: From single crystal to polycrystalline
  aggregates.
\newblock {\em International Journal of Plasticity}, 23(12):1957--1978, 2007.

\bibitem{essmann1979annihilation}
U~Essmann and HJPMA Mughrabi.
\newblock Annihilation of dislocations during tensile and cyclic deformation
  and limits of dislocation densities.
\newblock {\em Philosophical Magazine A}, 40(6):731--756, 1979.

\bibitem{castelluccio2017mesoscale}
Gustavo~M Castelluccio and David~L McDowell.
\newblock Mesoscale cyclic crystal plasticity with dislocation substructures.
\newblock {\em International Journal of Plasticity}, 98:1--26, 2017.

\bibitem{shenoy2008microstructure}
Mahesh Shenoy, Yustianto Tjiptowidjojo, and David McDowell.
\newblock Microstructure-sensitive modeling of polycrystalline in 100.
\newblock {\em International Journal of Plasticity}, 24(10):1694--1730, 2008.

\bibitem{armstrong1966mathematical}
Peter~J Armstrong, CO~Frederick, et~al.
\newblock {\em A mathematical representation of the multiaxial Bauschinger
  effect}, volume 731.
\newblock Berkeley Nuclear Laboratories Berkeley, CA, 1966.

\bibitem{zirkle2021micromechanical}
Theodore Zirkle, Ting Zhu, and David~L McDowell.
\newblock Micromechanical crystal plasticity back stress evolution within fcc
  dislocation substructure.
\newblock {\em International Journal of Plasticity}, 146:103082, 2021.

\bibitem{mcginty2001multiscale}
Robert~Davis McGinty.
\newblock {\em Multiscale representation of polycrystalline inelasticity}.
\newblock PhD thesis, Georgia Institute of Technology, 2001.

\bibitem{ling2005numerical}
Xianwu Ling, MF~Horstemeyer, and GP~Potirniche.
\newblock On the numerical implementation of 3d rate-dependent single crystal
  plasticity formulations.
\newblock {\em International Journal for Numerical Methods in Engineering},
  63(4):548--568, 2005.

\bibitem{mcginty2006semi}
RD~McGinty and DL~McDowell.
\newblock A semi-implicit integration scheme for rate independent finite
  crystal plasticity.
\newblock {\em International Journal of Plasticity}, 22(6):996--1025, 2006.

\bibitem{cuitino1993computational}
Alberto~M Cuitino and Michael Ortiz.
\newblock Computational modelling of single crystals.
\newblock {\em Modelling and Simulation in Materials Science and Engineering},
  1(3):225, 1993.

\bibitem{moose}
Plug-n-play system overview in tensor mechanics module in moose.
\newblock
  \url{https://mooseframework.inl.gov/modules/tensor_mechanics/plug_n_play.html}.
\newblock Accessed: 2022-11-12.

\bibitem{shi2022anisotropy}
Baodong Shi, Chong Yang, Yan Peng, Fucheng Zhang, and Fusheng Pan.
\newblock Anisotropy of wrought magnesium alloys: A focused overview.
\newblock {\em Journal of Magnesium and Alloys}, 2022.

\bibitem{agnew2001application}
SR~Agnew, MH~Yoo, and CN~Tome.
\newblock Application of texture simulation to understanding mechanical
  behavior of mg and solid solution alloys containing li or y.
\newblock {\em Acta Materialia}, 49(20):4277--4289, 2001.

\bibitem{kelley1968plane}
EW~Kelley and WFJR Hosford.
\newblock Plane-strain compression of magnesium and magnesium alloy crystals.
\newblock {\em Trans Met Soc AIME}, 242(1):5--13, 1968.

\bibitem{kelley1968deformation}
EW~Kelley.
\newblock The deformation characteristics of textured magnesium.
\newblock {\em Trans. of Metall. Soc. Of AIME}, 242:654--660, 1968.

\bibitem{simmons1971single}
Gene Simmons.
\newblock Single crystal elastic constants and caluculated aggregate
  properties.
\newblock {\em A handbook}, 4, 1971.

\bibitem{wu1996simulation}
PD~Wu, KW~Neale, and Erik Van~der Giessen.
\newblock Simulation of the behaviour of fcc polycrystals during reversed
  torsion.
\newblock {\em International Journal of Plasticity}, 12(9):1199--1219, 1996.

\bibitem{tanner1999deformation}
Albert~B Tanner and David~L McDowell.
\newblock Deformation, temperature and strain rate sequence experiments on ofhc
  cu.
\newblock {\em International Journal of Plasticity}, 15(4):375--399, 1999.

\bibitem{tanner1998thesis}
Albert~Buck Tanner.
\newblock {\em Modeling temperature and strain rate history effects in OFHC
  Cu}.
\newblock PhD thesis, Georgia Institute of Technology, 1998.

\bibitem{cardonne1995tantalum}
SM~Cardonne, P~Kumar, CA~Michaluk, and HD~Schwartz.
\newblock Tantalum and its alloys.
\newblock {\em International Journal of Refractory Metals and Hard Materials},
  13(4):187--194, 1995.

\bibitem{lee2023deformation}
Seunghyeon Lee, Hansohl Cho, Curt~A Bronkhorst, Reeju Pokharel, Donald~W Brown,
  Bj{\o}rn Clausen, Sven~C Vogel, Veronica Anghel, George~T Gray~III, and
  Jason~R Mayeur.
\newblock Deformation, dislocation evolution and the non-schmid effect in
  body-centered-cubic single-and polycrystal tantalum.
\newblock {\em International Journal of Plasticity}, page 103529, 2023.

\bibitem{hoge1977temperature}
KG~Hoge and AK~Mukherjee.
\newblock The temperature and strain rate dependence of the flow stress of
  tantalum.
\newblock {\em Journal of Materials Science}, 12(8):1666--1672, 1977.

\bibitem{cho2018anomalous}
Hansohl Cho, Curt~A Bronkhorst, Hashem~M Mourad, Jason~R Mayeur, and
  DJ~Luscher.
\newblock Anomalous plasticity of body-centered-cubic crystals with non-schmid
  effect.
\newblock {\em International Journal of Solids and Structures}, 139:138--149,
  2018.

\bibitem{nemat1997direct}
Sia Nemat-Nasser and JB~Isaacs.
\newblock Direct measurement of isothermal flow stress of metals at elevated
  temperatures and high strain rates with application to ta and taw alloys.
\newblock {\em Acta Materialia}, 45(3):907--919, 1997.

\bibitem{bingert1997texture}
JOHN~F Bingert, PB~Desch, SR~Bingert, PJ~Maudlin, and CN~Tom{\'e}.
\newblock Texture evolution in upset-forged p/m and wrought tantalum:
  experimentation and modeling.
\newblock Technical report, Los Alamos National Laboratory, 1997.

\bibitem{bratov2015comparison}
V~Bratov and EN~Borodin.
\newblock Comparison of dislocation density based approaches for prediction of
  defect structure evolution in aluminium and copper processed by ecap.
\newblock {\em Materials Science and Engineering: A}, 631:10--17, 2015.

\bibitem{lim2015quantitative}
Hojun Lim, Jay~D Carroll, Corbett~C Battaile, Brad~L Boyce, and Christopher~R
  Weinberger.
\newblock Quantitative comparison between experimental measurements and cp-fem
  predictions of plastic deformation in a tantalum oligocrystal.
\newblock {\em International Journal of Mechanical Sciences}, 92:98--108, 2015.

\end{thebibliography}

\end{document}